\documentclass[preprint,3p,times,twocolumn,authoryear]{elsarticle}
\usepackage{graphics}
\usepackage{amsmath}
\usepackage{amssymb}
\usepackage{multirow}
\usepackage{subfigure}

\usepackage[lined, linesnumbered, ruled]{algorithm2e}

\biboptions{comma,round}
\journal{Journal of Network and Computer applications}

\begin{document}

\begin{frontmatter}



\title{Agile-SD: A Linux-based TCP Congestion Control Algorithm for Supporting High-speed and Short-distance Networks}
% This paper has been submitted on October 21, 2014 then it has been accepted with revision on March 01, 2015.
% It has been resubmitted after the first revision on March 13, 2015.


\author[lable1]{Mohamed A. Alrshah$^{1,}$}
\author[lable1]{Mohamed Othman$^{2,}$}
\author[lable2]{Borhanuddin Ali}
\author[lable1]{Zurina Mohd Hanapi}
\address[lable1]{Department of Communication Technology and Network, Universiti Putra Malaysia, 43400 UPM, Serdang, Selangor D.E., Malaysia}
\address[lable2]{Department of Computer and Communication Systems Engineering, Universiti Putra Malaysia, 43400 UPM, Serdang, Selangor D.E, Malaysia.}

\begin{abstract}
Recently, high-speed and short-distance networks are widely deployed and their necessity is rapidly increasing everyday. This type of networks is used in several network applications; such as Local Area Networks (LAN) and Data Center Networks (DCN). In LANs and DCNs, high-speed and short-distance networks are commonly deployed to connect between computing and storage elements in order to provide rapid services. Indeed, the overall performance of such networks is significantly influenced by the Congestion Control Algorithm (CCA) which suffers from the problem of bandwidth under-utilization, especially if the applied buffer regime is very small. In this paper, a novel loss-based CCA tailored for high-speed and Short-Distance (SD) networks, namely Agile-SD, has been proposed. The main contribution of the proposed CCA is to implement the mechanism of agility factor. Further, intensive simulation experiments have been carried out to evaluate the performance of Agile-SD compared to Compound and Cubic which are the default CCAs of the most commonly used operating systems. The results of the simulation experiments show that the proposed CCA outperforms the compared CCAs in terms of average throughput, loss ratio and fairness, especially when a small buffer is applied. Moreover, Agile-SD shows lower sensitivity to the buffer size change and packet error rate variation which increases its efficiency.

\end{abstract}


\end{frontmatter}

\footnotetext[1]{Corresponding authors: 
\\E-mail addresses: mohamed.asnd@gmail.com (Mohamed Alrshah),\\mothman@upm.edu.my (Mohamed Othman).}

\footnotetext[2]{The author is also an associate researcher at the Computational \mbox{Science} and Mathematical Physics Lab, Institute of Mathematical \mbox{Science}, Universiti Putra Malaysia.}

%\linenumbers

%**********************************************************************************************************************************************************************************
%\underline{Important note:} The small value of delay can disturb the performance of delay-based TCP variants because most of the deployed delay-based TCPs are reading the rounded delay from the kernel as milliseconds which hide the variation of last obtained $RTT$ from the $RTT_{base}$.  
%*********************************************************************************************************************************************************************************

\section{Introduction}
\label{Intro}
\begin{figure} [t]
\centering
\includegraphics[width=1\linewidth]{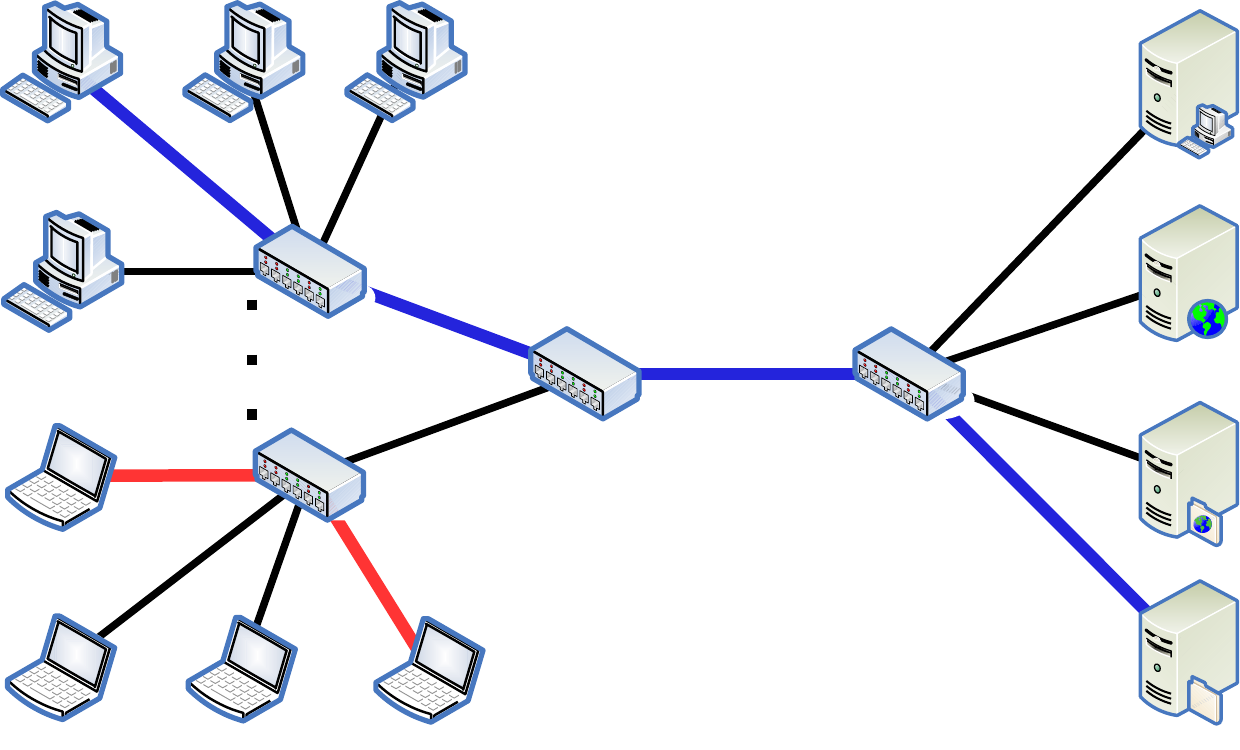}
\caption{A Network Topology for LAN.}
\label{fig:lantopology}
\end{figure}

\begin{figure}[t]
\centering
\includegraphics[width=1\linewidth]{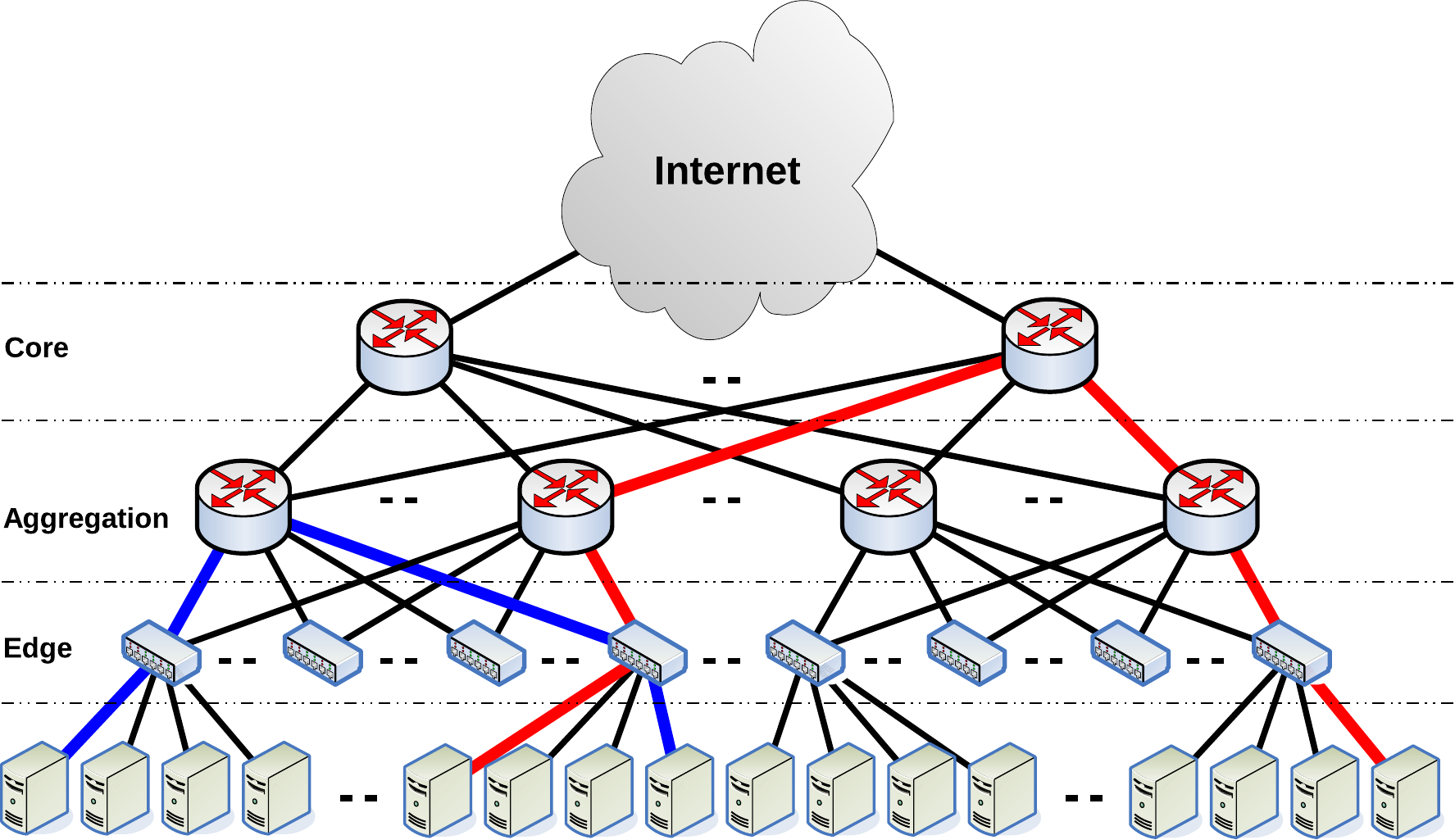}
\caption{Multi-rooted Hierarchical Topology of Data Centers.}
\label{fig:dctopology}
\end{figure}

In the last decades, the necessity of high-speed and short-distance networks is rapidly increasing everyday due to their wide deployment. Several network applications, such as Local Area Networks (LAN) and Data Center Networks (DCN), are implementing this type of networks \citep{Buyya2008, Armbrust2010}. These LANs and DCNs serve a very wide range of network-based applications; such as web hosting, searching engines, social media, multimedia broadcasting and storage drives. In the environment of LANs, as shown in Figure \ref{fig:lantopology}, and DCNs, as shown in Figure \ref{fig:dctopology} \citep{Alfares2010, Wu2012, Yoo2012, prakash2012}, high-speed and short-distance networks are commonly deployed to connect computing and storage elements to each other in order to provide rapid services.These networks have certain characteristics which are widely different from other types of networks; for instance, link delay is very small which can be a few milliseconds or even hundreds of microseconds and the Bandwidth-Delay-Product (BDP) of the link is very small compared to its equivalent in high-speed and long-distance networks \citep{Tahiliani2012, Vasudevan2009}.

These attributes could negatively affect the performance of the Transmission Control Protocol (TCP) by making it either more aggressive or more conservative based on the applied approach. In fact, the Congestion Control Algorithm (CCA) is one of the main parts of TCP. It significantly affects the overall performance of such networks, because it is still suffering from the problem of bandwidth under-utilization, especially if the applied buffer regime is very small. This under-utilization of bandwidth is caused by the variation of the aforementioned characteristics of the networks which results either a slow growth of $cwnd$ or an over-injection of data into the network \citep{Afanasyev2010, Scharf2011, Callegari2012b, Callegari2014, Lar2013, acharya2012, alrshah2014}.

In order to solve the problem of bandwidth under-utilization over high-speed and Short-Distance (SD) networks, a new loss-based CCA, namely Agile-SD, has been proposed. The main contribution of the proposed CCA is to implement the mechanism of agility factor. Further, intensive simulation experiments have been carried out to evaluate the performance of Agile-SD compared to Compound (the default CCA of MS Windows since Windows Vista) and Cubic (the default CCA of Linux since Kernel 2.6.16) which are the default CCAs of the most commonly used operating systems \citep{Afanasyev2010, alrshah2014}.

The rest of this paper is organized as follows: the related work is presented in Section \ref{RW} while Section \ref{Agile-SD} presents the proposed algorithm \textquotedblleft Agile-SD\textquotedblright. Section \ref{PE} explains the used approach of performance evaluation which is contains the experiments' setup, network topology, performance metrics, results and discussion. Finally, Section \ref{Conc} presents the conclusion and the future work.

\section{Related Work}
\label{RW}

In order to solve the problem of bandwidth under-utilization, Cubic \citep{Ha2008}, Scalable TCP \citep{Kelly2003}, HS-TCP \citep{Floyd2003}, BIC \citep{xu2004}, HCC \citep{xu2011}, H-TCP \citep{Leith2004}, TCP Africa \citep{King2005}, TCP Compound \citep{Tan2006}, Fusion \citep{Kaneko2007}, TCP illinois \citep{Liu2008} and YeAH \citep{Baiocchi2007} have been developed and implemented in the real operating systems. All of these TCP variants are still unable to fully utilize the available bandwidths of high-speed networks, especially if the used buffer size is less than the BDP of the link \citep{Afanasyev2010, Scharf2011, Callegari2012b, Callegari2014, Lar2013, acharya2012, alrshah2014}. 

Further, some researchers tried to solve the aforementioned problem by proposing a set of CCAs or TCP variants; such as DCTCP \citep{Alizadeh2010}, ICTCP \citep{Haitao2013}, IA-TCP \citep{Jaehyun2012} and D$^2$TCP \citep{Vamanan2012} which designed for data center networks. All of these TCP variants are still suffering from some critical problems, such as the problem of TCP outcast which has not been solved yet \citep{Tahiliani2012}.

Furthermore, some researchers tried to improve the performance of TCP by using parallel approaches; such as AppTCP \citep{Wang2013}, GridFTP \citep{allcock2005}, pTCP \citep{hsieh2002}, BBCP \citep{hanushevsky2001}, PSockets \citep{sivakumar2000}, MulTCP \citep{crowcroft1998}, DPSS \citep{Tierney1994} and Parallel-TCP \citep{alrshah2009, alrshah2013}. Most of parallel schemes have achieved high bandwidth utilization, but unfortunately they have another issues which limited their deployments. One of these issues is that all of the parallel schemes have a very high aggressiveness level compared to the existing single-based TCP variants. This aggressive behavior negatively affects the fairness \citep{Fu2005, fu2007}.

However, the widely deployed CCAs; such as Compound and Cubic which have been set as the default CCAs in MS Windows and Linux operating systems, respectively; are playing the important role in the real networks. Thus, Compound and Cubic should be used as a benchmark, to confirm the performance of any newly proposed CCA or TCP variant. For this reason, these two CCAs are going to be briefly explained in the next subsections.

\subsection{Compound TCP (C-TCP)}
The widely deployed TCP variant namely C-TCP \citep{Tan2006} is the default CCA of MS Windows since Windows Vista. C-TCP combines HS-TCP \citep{Floyd2003} and NewReno \citep{floyd1999} to be used as fast and slow modes, respectively. C-TCP is a loss-delay-based approach relies on multi-modes switching to increase the bandwidth utilization over high-speed networks. Generally, it improves the performance of TCP to some extent but it introduces another problem which is the RTT mis-estimation. This problem has been inherited from Vegas \citep{brak1995} and it can negatively affect the overall performance of the protocol \citep{Afanasyev2010, alrshah2014}.

\subsection{CUBIC TCP (Cubic)}
Cubic \citep{Ha2008} is the default CCA of Linux operating systems since its implementation in Kernel 2.6.16. Cubic enhances the bandwidth utilization over high-speed networks by increasing the $cwnd$ in the congestion avoidance phase by a $cubic$ function of the elapsed time since last loss. In addition, Cubic forces its $cwnd$ not to be less than the pre-calculated $cwnd$ of NewReno. Despite of all, Cubic is still suffering from the under-utilization of high-speed bandwidth specifically when the used buffer size is small \citep{Afanasyev2010, alrshah2014, Ha2008}.

\subsection{The Latest Issues}
Recent studies have revealed that all of the current TCP variants have different levels of inability on fully utilizing the bandwidths over the new generation of high-speed networks, especially if a \emph{near-zero} buffer is applied. Thus, it becomes very necessary to design a new CCA to increase the bandwidth utilization over such networks \citep{Afanasyev2010, alrshah2014}.

%\section{Related Work}
%\label{LR}

\section{Agile-SD: The Proposed Algorithm}
\label{Agile-SD}

\begin{figure}[h!]
\centering
\includegraphics[width=0.7\linewidth]{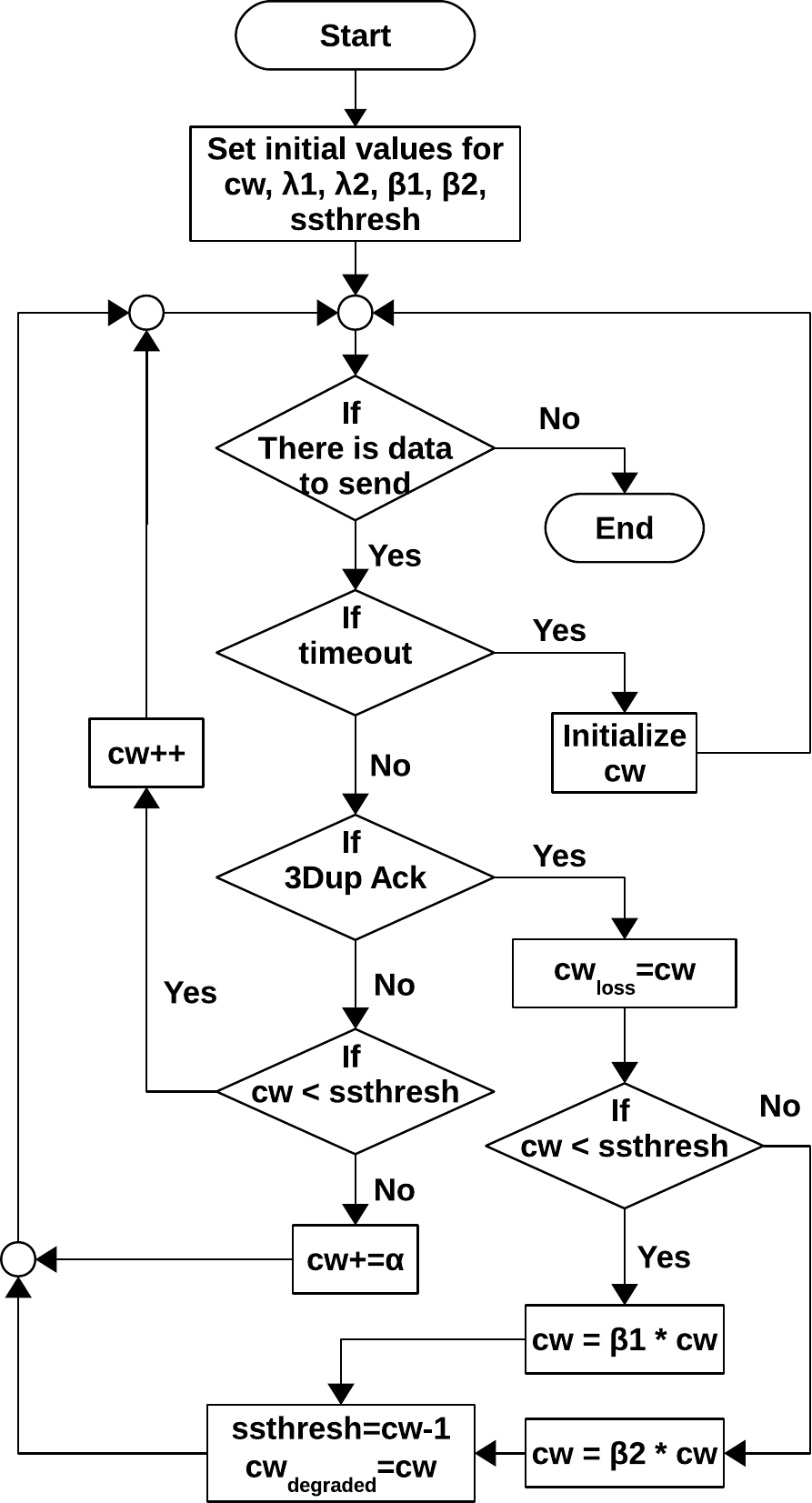}
\caption{The flow control diagram of Agile-SD.}
\label{fig:TCP}
\end{figure}

Algorithm \ref{algo01} explains the Agile-SD mechanism which is geared to work on high-speed and short-distance networks to enhance the overall performance and bandwidth utilization while preserving the fairness. Moreover, Figure \ref{fig:TCP} shows the flow control diagram of Agile-SD and the following subsections explain the proposed algorithm in more details.

%*********************Algorithm Keywords Definitions*************%
\SetKwProg{Function}{Function}{ }{end}
\SetKwProg{Event}{Event}{ do}{end}
%****************************************************************%
\begin{algorithm}[t!]
\caption{Agile-SD Congestion Avoidance.}\label{algo01}

\textbf{Initialization:}\\
%\hspace{1cm}$set\_params()$\\
	\hspace{0.5cm}$\lambda_{min} \leftarrow 1, \lambda_{max} \leftarrow 3,$\\
	\hspace{0.5cm}$\beta_1 \leftarrow 0.90, \beta_2 \leftarrow 0.95,$\\
	\hspace{0.5cm}$cwnd \leftarrow 2$

\Event{On ACK Receiption}
{
	calculate $gap_{current}$ as in Equation (\ref{eq3})\\
	
	calculate $gap_{total}$ as in Equation (\ref{eq2})\\

	calculate $\lambda$ as in Equation (\ref{eq1})\\

	$\alpha = \frac{\lambda}{cwnd}$
	%calculate $\alpha$ as in Equation (\ref{eq1})\\
	
	$cwnd \leftarrow cwnd + \alpha$\\
}

%\Event{On Three Duplicate ACKs Detection}
\Event{On Loss Detection of 3-duplicated ACKs}
{
	$cwnd_{loss} 	\leftarrow cwnd$\\
	
	\uIf {$tcp\_status = SlowStart$}
	{	
		$cwnd \leftarrow cwnd \times \beta_1$ \label{L13}\\
	}
	\Else
	{
		$cwnd \leftarrow cwnd \times \beta_2$ \label{L17}\\
	}
	
	$ssthresh 	\leftarrow cwnd - 1$\\
	$cwnd_{degraded} \leftarrow cwnd$\\
}
\end{algorithm} \DecMargin{1em}
%****************************************************************%

\subsection{The Agility Factor Mechanism}
As known, \citet{RFC6928} increases the initial value of TCP $cwnd$ to 10 packets, but Agile-SD initializes its $cwnd$ by 2 packets in order to focus on the impact of CA on bandwidth utilization. However, in the future implementations the initial $cwnd$ should be set to 10 to gain better bandwidth utilization.

Clearly, Agile-SD increases its $cwnd$ in the stage of congestion avoidance by fraction similarly as the existing CCAs. But, Agile-SD increases its $cwnd$ by $\frac{\lambda}{cwnd}$ to show a convex curve unlike the standard TCP which increases its $cwnd$ linearly by $\frac{1}{cwnd}$. The main contribution of Agile-SD is the unique $cwnd$ growth function which relies on the agility factor mechanism which symbolized by $\lambda$, as shown in Equation \eqref{eq1}.
\begin{figure}[t]
\centering
\includegraphics[width=\linewidth]{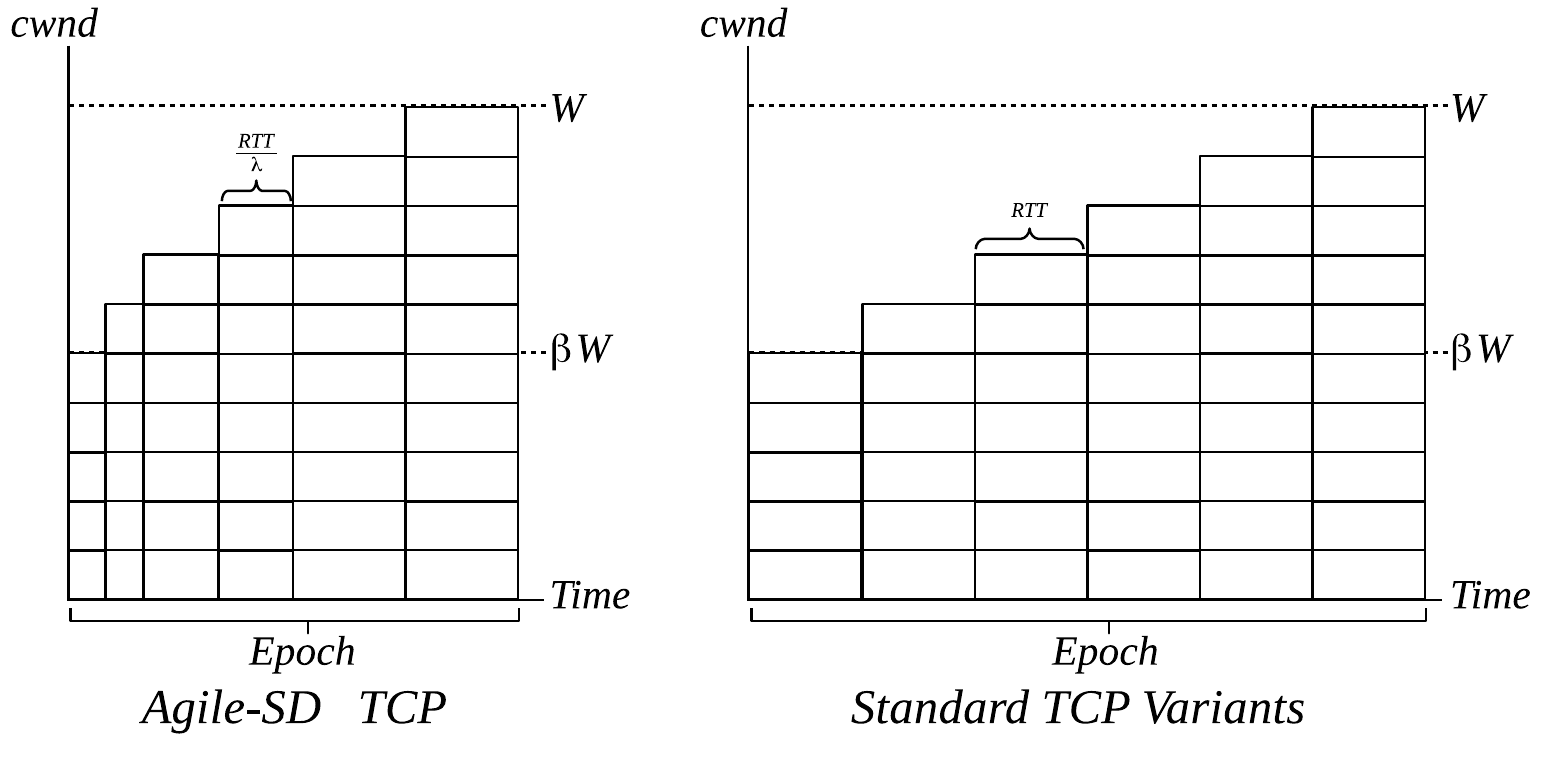}
\caption{The $cwnd$ evolution of Agile-SD and the standard TCP.}
\label{fig:epoch}
\end{figure}
\begin{align}
&\lambda = max\left( \frac{\lambda_{max} \times gap_{current}}{gap_{total}}, \lambda_{min}\right)	\label{eq1}
\end{align}
where, $gap_{total}$ exemplifies the amount of decrease in the $cwnd$ which caused by the loss event. In other words, $gap_{total}$ represents the released amount from the bandwidth after loss. $gap_{total}$ is calculated as the distance between the maximum recorded limit of the bandwidth ($cwnd_{loss}$) and the $cwnd_{degraded}$ as in Equation \eqref{eq2}.
\begin{align}
gap_{total} = max\left((cwnd_{loss} - cwnd_{degraded}), 1\right)							\label{eq2}
\end{align}

While, $gap_{current}$ is calculated as the difference between the maximum recorded limit of the bandwidth ($cwnd_{loss}$) and the current $cwnd$, as in Equation \eqref{eq3}.
\begin{align}
gap_{current} = max\left((cwnd_{loss} - cwnd), 1\right)										\label{eq3}
\end{align}

Simply, $\lambda$ is used to mitigate the impact of loss degradation on the overall performance of TCP. Specifically, $\lambda$ shortens the time of epoch which is needed by CCA to move its $cwnd$ from $cwnd_{degraded}$ to $cwnd_{loss}$ or to the maximum allowed $cwnd$, as shown in Figure \ref{fig:epoch}. In order to increase the bandwidth utilization, $\lambda$ speeds up the growth of $cwnd$ when the current $cwnd$ is far away from the last $cwnd_{loss}$. While, it conservatively slows down the growth of $cwnd$ when the current $cwnd$ nearing to the last $cwnd_{loss}$.

More specifically, to ensure that the performance of Agile-SD is not less than the standard TCP, $\lambda_{min}$ must be always set to 1 while $\lambda_{max}$ must be always set to a value $\geq 1$. However, if $\lambda_{max}$ is set to 1, Agile-SD will behave exactly similar to NewReno. But, if it was set to a value $> 1$, such as 2, 3 or 4, it would clearly improve the overall performance. 

\begin{figure} [t]
\centering
\includegraphics[width=0.9\linewidth]{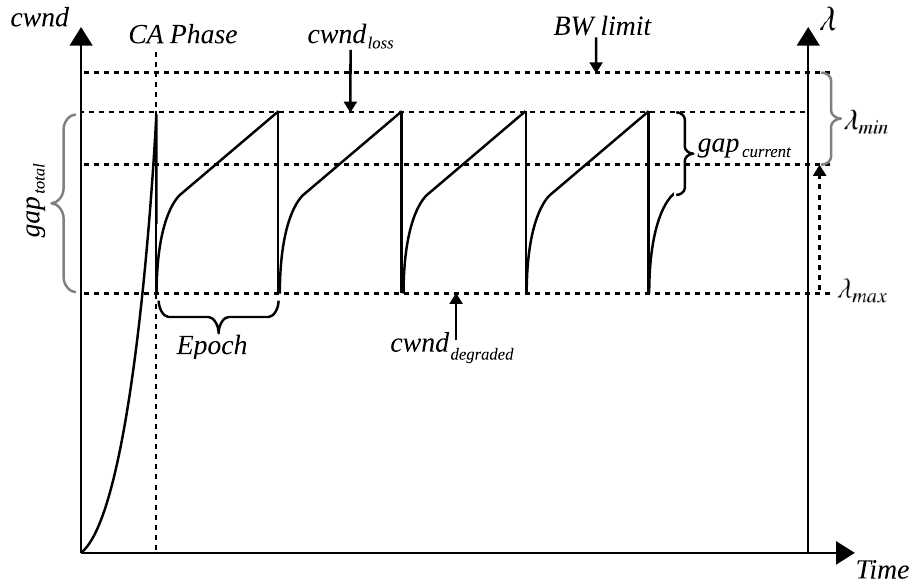}
\caption{The concept of agility factor mechanism $\lambda$.}
\label{fig:factor}
\end{figure}

\begin{figure} [t]
\centering
\includegraphics[width=0.9\linewidth]{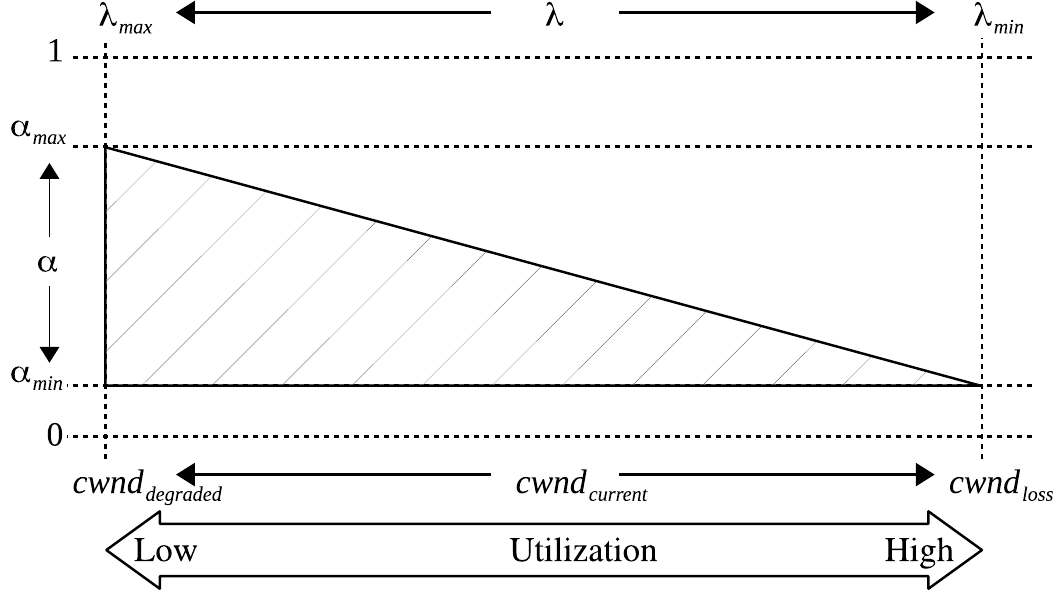}
\caption{Relation between $\alpha, \lambda$ and the bandwidth utilization.}
\label{fig:fraction}
\end{figure}

After loss detection by receiving three duplicate ACKs, Agile-SD initiates its agility factor by $\lambda_{max}$, then reduces it every cycle towards $\lambda_{min}$. Nevertheless, it restarts by $\lambda_{max}$ again if another three duplicate ACKs are received as shown in Figure \ref{fig:factor}. Consequently, it is very clear that $\alpha$ is directly proportional to $\lambda$, and $\lambda$ is reversely proportional to the current size of $cwnd$, as shown in Figure \ref{fig:fraction}. In other words, the aggressiveness of the proposed algorithm increases whenever the difference between $cwnd$ and $cwnd_{loss}$ increases, as in Equation \eqref{eq4}. 

\begin{align}
\alpha_{max} &= \lim_{gap_{current} \to gap_{total}}  \dfrac{max\left( \frac{\lambda_{max} \times gap_{current}}{gap_{total}}, \lambda_{min}\right)} {cwnd}\label{eq4}\\
&= \frac{\lambda_{max}}{cwnd} \nonumber
\end{align}

Contrarily, it decreases its aggressiveness whenever the utilization is around to touch the maximum available bandwidth, as in Equation \eqref{eq5}. 
\begin{align}
\alpha_{min} &= \lim_{gap_{current} \to zero}  \dfrac{max\left( \frac{\lambda_{max} \times gap_{current}}{gap_{total}}, \lambda_{min}\right)} {cwnd}\label{eq5}\\
&= \frac{\lambda_{min}}{cwnd} \nonumber
\end{align}

In general, Agile-SD reduces the cycle time which by its role reduces the epoch time to overcome the problem of slow evolution of $cwnd$ provided by the standard TCP, as shown early in Figure \ref{fig:epoch}. On one hand, this behavior guarantees the performance of Agile-SD not to be lower than the standard TCP. Thus, it increases the bandwidth utilization by improving the ability of Agile-SD to expose the condition of the underlying network as shown in Figure \ref{fig:fraction}. On the other hand, it reduces the sensitivity of Agile-SD to the loss rate.

\subsection{The Decrement of cwnd}

The standard TCP applies the \textit{Multiplicative} \textit{Decrease} mechanism which halves the $cwnd$ after any loss detection, by receiving three duplicate ACKs, regardless of which stage the loss is detected in. Unlikely, Agile-SD decreases its $cwnd$ after any loss detection by two ways based on the stage which the loss is coming from. First, if the loss is detected in the slow start stage, Agile-SD reduces its $cwnd$ to $\beta_1\%$ of the latest $cwnd$ as shown at Line \ref{L13} in Algorithm \ref{algo01}. Second, if the loss is detected in the congestion avoidance stage, Agile-SD reduces its $cwnd$ to $\beta_2\%$ of the latest $cwnd$  as shown at Line \ref{L17} in Algorithm \ref{algo01}. Moreover, Agile-SD sets the $ssthresh$ to $cwnd - 1$, after any degradation, in order to avoid slipping into an undesirable slow start.

Since, the loss which happens in the slow start stage is more severe than which happens in the congestion avoidance stage. Therefore, the value of $\beta_1$ should be always less than $\beta_2$. In other words, the reduction which follows a slow start loss should be greater than the reduction which follows a congestion avoidance loss. Also, $\beta_1$ and $\beta_2$ must be reversely proportional to their relative $\lambda_{max}$ but the relation among them is still under investigation to be revealed in the future work. Thus, whenever the values of $\beta_1$ and $\beta_2$ is increased, the value of $\lambda_{max}$ should be adaptively decreased and vice versa. For instance, if $\beta_1$ and $\beta_2$ are set to 0.9 and 0.95, respectively, where these values are compatible with $\lambda_{max} = 3$. Then, if $\beta_1$ and $\beta_2$ are reduced to 0.85 and 0.9, respectively, the value of $\lambda_{max}$ should be increased to a value, such as 4 or 5, and so on.

\subsection{Agile-SD Overall Behavior}

Similar to standard TCP, Agile-SD starts by slow start to show an exponential increase until the detection of first loss; by receiving three duplicate ACKs. This reduces its $cwnd$ to $\beta_1\%$ and triggers the congestion avoidance function. In this stage, Agile-SD increases its $cwnd$ by $\alpha$ to show a convex curve. But, if the $cwnd$ becomes closer to the bandwidth limit which is set as $cwnd_{loss}$, it starts a linear increase until detecting another packet loss. If the event of packet loss is detected, Agile-SD reduces its $cwnd$ to $\beta_2\%$ then repeats the same stages which follows the slow start stage. However, if timeout is detected at any stage, Agile-SD resets its $cwnd$ to the initial value, as shown in Figure \ref{fig:TCP}.

For more understanding, assume that there is a TCP link with $cwnd_{loss} = 12, cwnd_{degraded} = 9$ and a constant $RTT$ equal to 20 ms, and the congestion avoidance stage is just started after the loss directly. Thus, the number of $cycles$ needed by any CCA to reach $cwnd_{loss}$ is 4 cycles which is equal to $(cwnd_{loss} - cwnd_{degraded} + 1)$. Consequently, the epoch time needed by the standard TCP, which is $"RTT-dependent"$, is the number of needed $cycles$ times $RTT$, so it will be equal to 80 ms. 

Instead, Agile-SD increases its $cwnd$ independently from the $RTT$. Thus, every $cycle$ consumes a time of $\frac{RTT}{\lambda}$ to send a number of $\frac{cwnd}{\lambda}$ packets during that cycle, then it increases its $cwnd$ by 1. Consequently, the epoch time needed by Agile-SD will be equal to what shown in Equation \eqref{eq6}.
\begin{align}
&Epoch Time = \sum_{i=1}^{k}\dfrac{RTT}{\lambda_i}  \label{eq6}
%\\ \text{where } k &\text{ is the number of needed cycles}. \nonumber
\end{align}
where $k$ is the number of needed cycles. 

Suppose $\lambda_{min}$ and $\lambda_{max}$ are set to 1 and 4, respectively. So, $\lambda_i$ will take the value of [4, 3, 2, 1] sequentially, which will result in an epoch time equal to 41.66 ms. Thus, the epoch time of Agile-SD will be shrunk by around 48\% from the epoch time of the standard TCP on the same network link. This behavior helps Agile-SD to increase its $cwnd$ more quickly than the other compared CCAs and consequently improves the bandwidth utilization. In other words, the faster $cwnd$ growth is the higher bandwidth utilization and vice versa. 

\section{Performance Evaluation of Agile-SD}
\label{PE}
The goal of this work is, to develop a new CCA, namely Agile-SD, which has the ability of increasing the bandwidth utilization over high-speed and short-distance networks while maintaining fairness. \mbox{Agile-SD} CCA has been implemented as a pluggable Linux CCA module which can be plugged into any Linux Kernel. As well as, this module has the ability to be plugged into NS-2 network simulator, as a Linux TCP, in order to evaluate its performance compared to some of the widely deployed CCAs.

\subsection{The Experiments Setup}
In this work, intensive simulation experiments have been conducted using the well-known network simulator NS-2 version 2.35, to evaluate the proposed CCA by comparing its performance with C-TCP and \mbox{Cubic}. The conducted experiments have been divided into three main scenarios: single-flow, sequentially established/terminated multiple-flows, and synchronously established/terminated multiple-flows. Table \ref{params} shows the setting of the experiments' parameters as used in this work.
\begin{table}[h!]
	\caption{Experiment Parameters.}
	\begin{center}
	\begin{tabular}{p{0.15cm}p{2.24cm}p{4.16cm}} %{lll} %
	\hline
	No.& Parameter			&	Value									\\ \hline
	1. & CCAs				&	Agile-SD, Cubic, C-TCP					\\ %%\hline
	2. & Link capacity		&	1 Gbps for all							\\ %%\hline
	3. & Link delay			&	1ms (node to router)						\\ %%\hline
	   & 					&	4ms (router to router)					\\ %%\hline
	4. & BDP				&	750KB (As in \citep{RFC1072})			\\ %%\hline
	5. & PER				&   $zero, 10^{-5}, 10^{-4}$				\\%%\hline
	6. & Buffer size		&	from 5 to 500 packets					\\ %%\hline
	7. & Packet size		&	1000 bytes								\\ %%\hline
	8. & Queuing Algo	 	&	Drop-Tail								\\ %%\hline	
	9. & Traffic type		&	FTP										\\ %%\hline
	10. & SACK, FACK		&	Disabled								\\ %%\hline
	11. & Simulation time	&	100 seconds								\\ \hline
	\end{tabular}
	\label{params}
	\end{center}
\end{table}

\begin{figure} [t!]
\centering
\includegraphics[width=\linewidth]{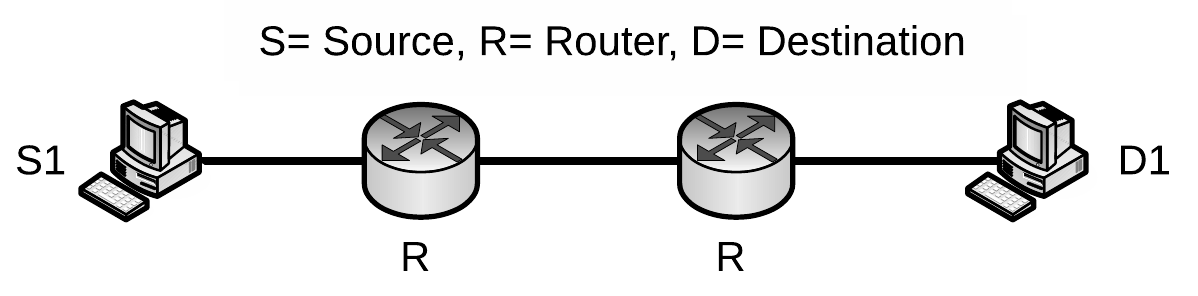}
\caption{Non-congested Network topology.}
\label{fig:topology-ideal}
\end{figure}

\begin{figure} [t!]
\centering
\includegraphics[width=\linewidth]{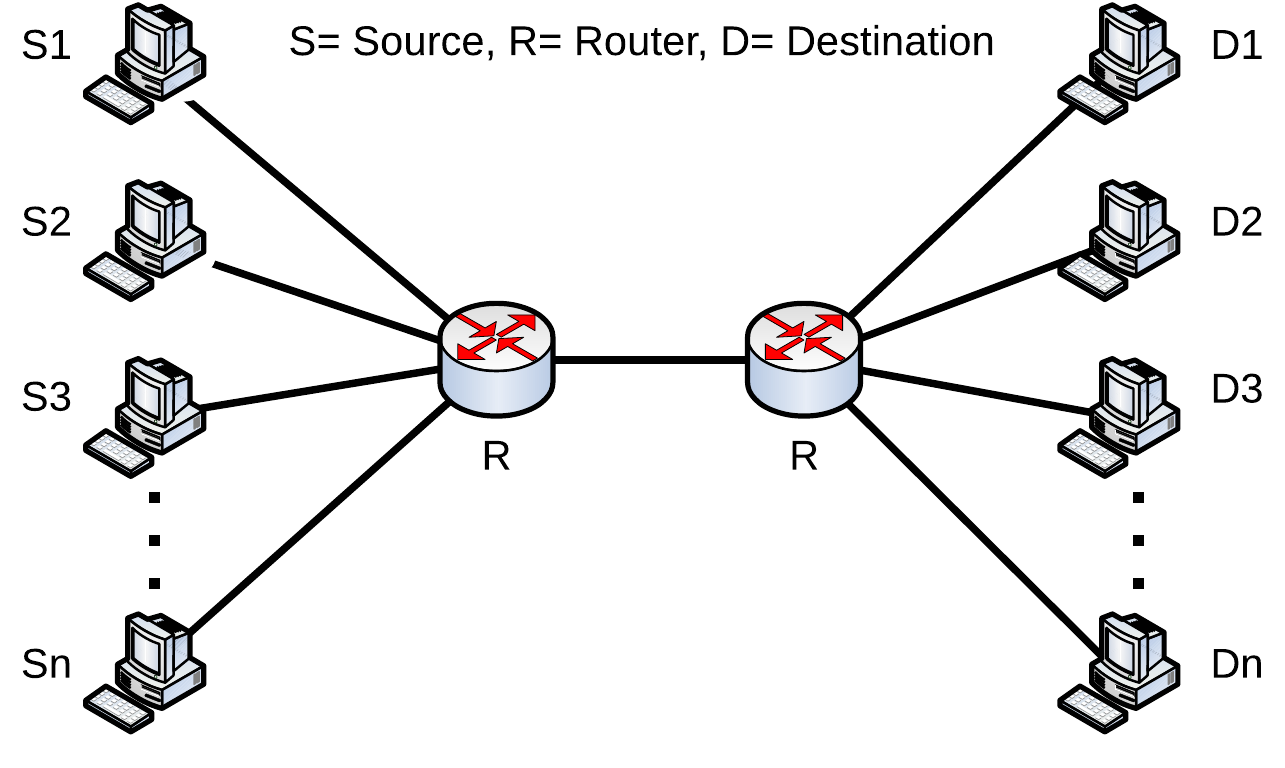}
\caption{Network topology with standard dumbbell bottleneck.}
\label{fig:topology}
\end{figure}

\begin{figure} [t!]
	\centering
	\begin{center}
		\subfigure[Sequentially established/terminated flows scenario.] 
		{
			\includegraphics[width=0.9\linewidth]{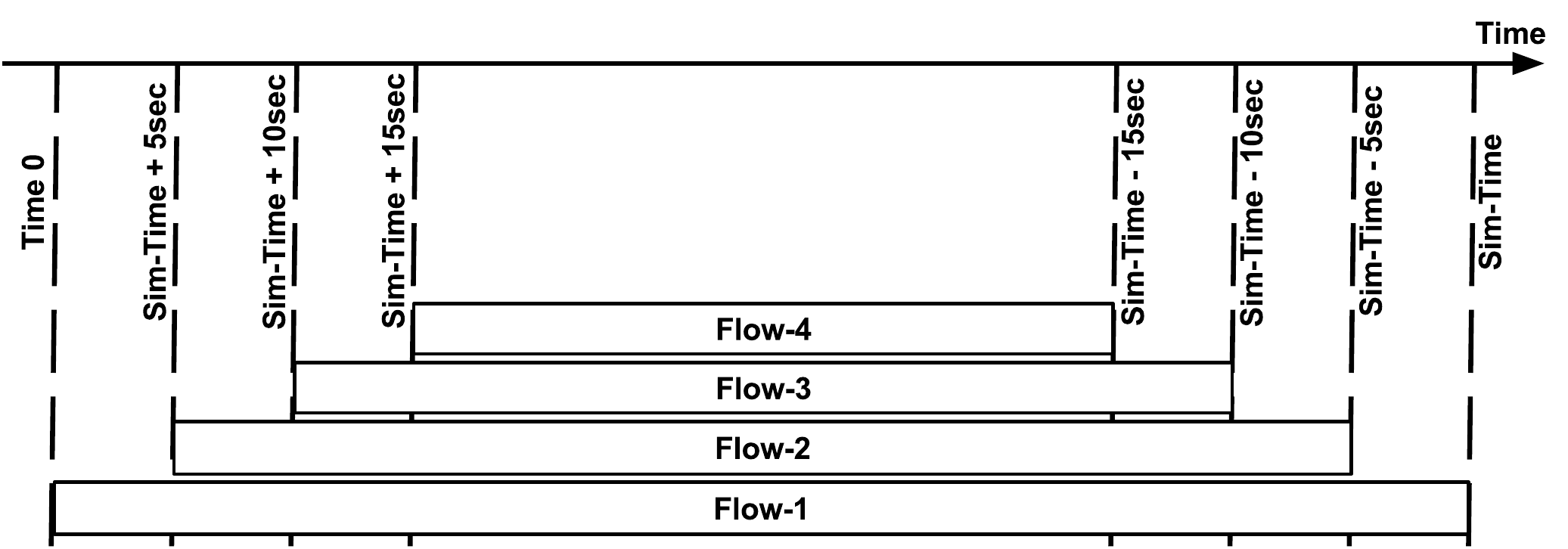}
			\label{fig:multi-flows-sequence1}
		}
		\subfigure[Synchronously established/terminated flows scenario.] 
		{
			\includegraphics[width=0.9\linewidth]{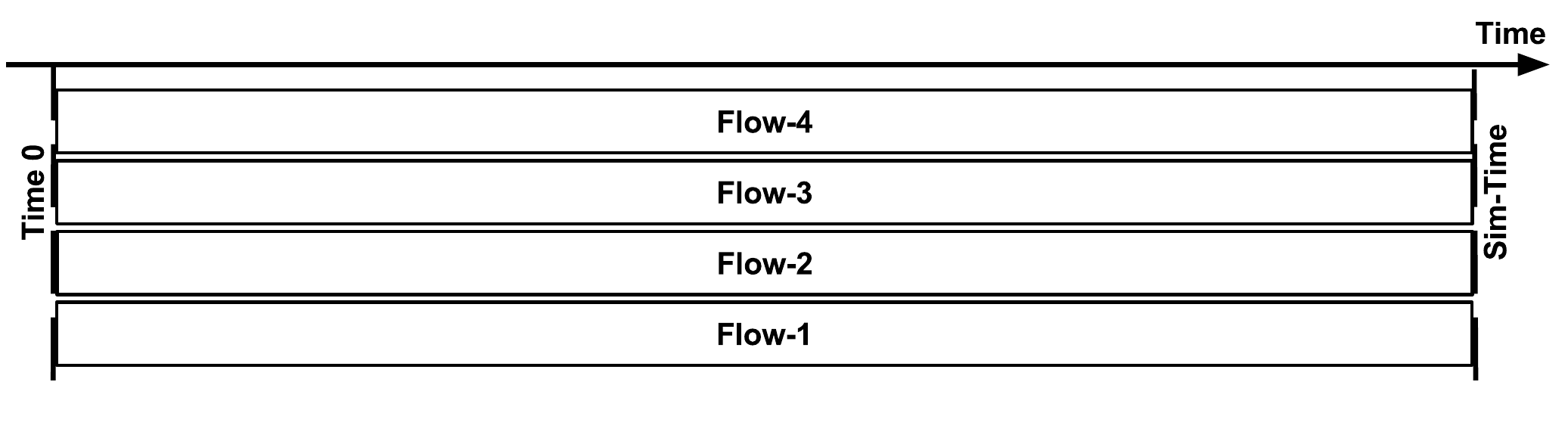}
			\label{fig:multi-flows-sequence2}
		}
	\end{center}
	\caption{The sequence of establishments and terminations of the multiple flows scenarios.}
	\label{fig:multi-flows-sequence}
\end{figure}

In the first scenario of single-flow, there is only one pair of sender and receiver, as shown in Figure \ref{fig:topology-ideal}, which presents an ideal case with no congestion to show the ability of the evaluated CCAs on achieving full bandwidth utilization. As for the second and third scenarios of multiple flows, there are $n$ pairs of sender and receiver, as shown in Figure \ref{fig:topology}, which have been used to simulate the network congestion and to show its impact on the performance measurements of the evaluated CCAs. In the second scenario, the flows are sequentially established and terminated as shown in Figure \ref{fig:multi-flows-sequence}(a). While in the third scenario, the flows are synchronously established and terminated as shown in Figure \ref{fig:multi-flows-sequence}(b).

In all of these experiments, a standard single dumbbell topology has been used, as shown in figures \ref{fig:topology-ideal} and \ref{fig:topology}, where $n$ is the competing senders ($S1$, $S2$, $S3$, ..., $Sn$) which send data simultaneously to $n$ receivers ($D1$, $D2$, $D3$, ..., $Dn$) through the shared single bottleneck. All source and destination nodes are connected to the bottleneck routers over LAN with $1Gbps$ speed and $1ms$ propagation delay. While the bottleneck link is $1Gbps$ speed with a propagation delay of $4ms$ \citep{Wang2013}. These experiments have been repeated for each CCA separately with variable buffer size and variable packet error rate (PER). The buffer size varies from $5$ to $500$ packets while the PERs which have been used are $10^{-4}$, $10^{-5}$ and $zero$ PER. 

Moreover, the performance metrics evaluated in this paper are the average throughput, loss ratio, inter-fairness, intra-fairness and RTT-fairness. Average throughput and loss ratio are evaluated to reflect the ability of the TCP variant on utilizing the bandwidth. While, measuring inter-fairness, intra-fairness and RTT-fairness is to show the quality of sharing the link between the competing TCP flows based on Jain's fairness index (JFI) \citep{jain1984}, as shown in Equation \eqref{fair}.

\begin{align}
JFI(x_1, x_2, ..., x_n) = \dfrac{(\sum_{i=1}^{n} x_i)^2}{n \cdot \sum_{i=1}^{n} x_i^2} 								\label{fair}
\end{align}

Substantially, these experiments show the impact of bottleneck congestion, buffer size and PER on the performance of the examined CCAs and also show the performance changes when a smaller buffer size is applied. Moreover, the simulation time used in all experiments has been set to 100 seconds which is enough for TCP to show its steady state.

\subsection{Results and Discussion}

This subsection presents an analytical discussion of the behavior exhibited by the proposed CCA and the compared CCAs. As well as, it presents the results of the performance evaluation and shows the measurements of the average throughput, loss ratio, inter-fairness, intra-fairness and RTT-fairness.

\subsubsection{The $cwnd$ evolution}
\begin{figure*} [t!]
	\centering
	\begin{center}
		\subfigure [Buffer Size = 5 Packets.]	{
			\includegraphics[scale=0.39]{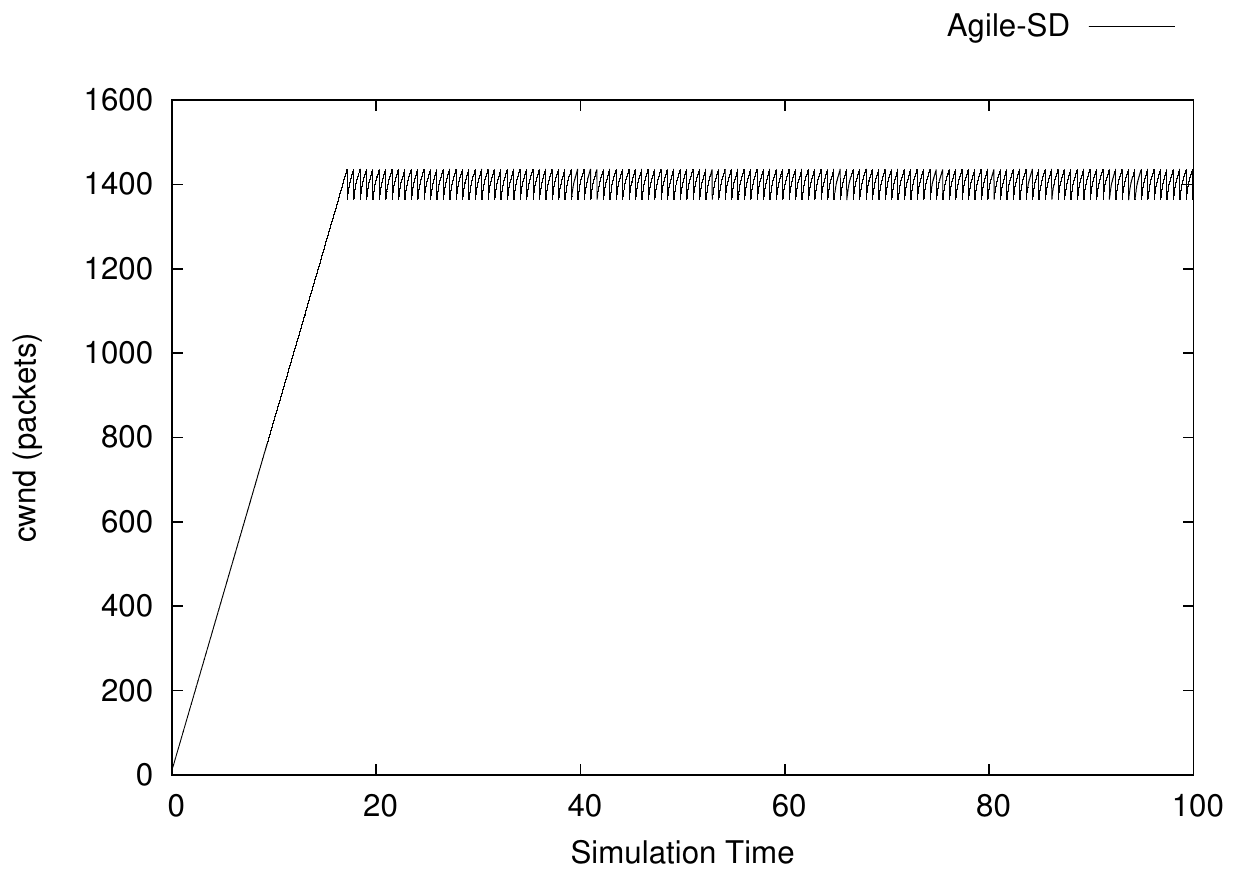}		
			\includegraphics[scale=0.39]{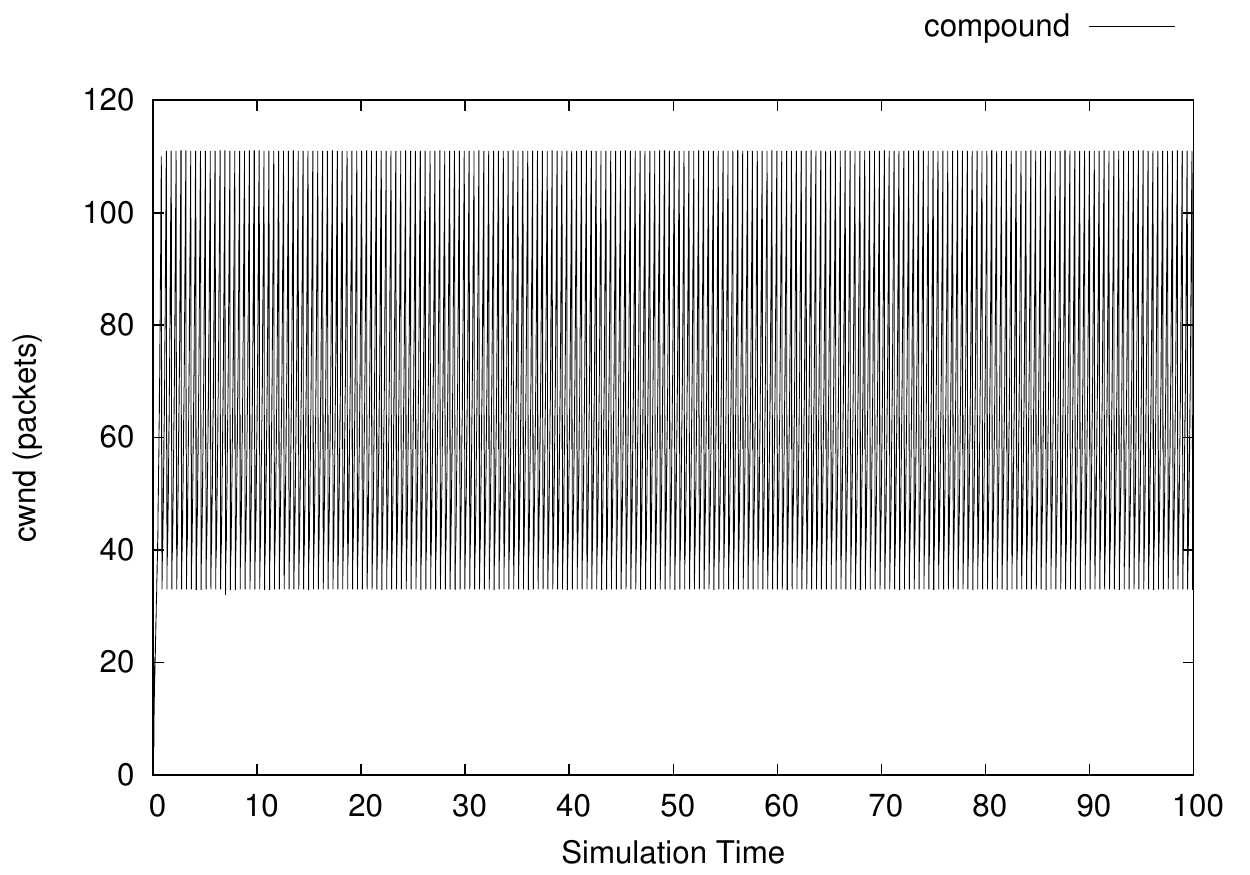}
			\includegraphics[scale=0.39]{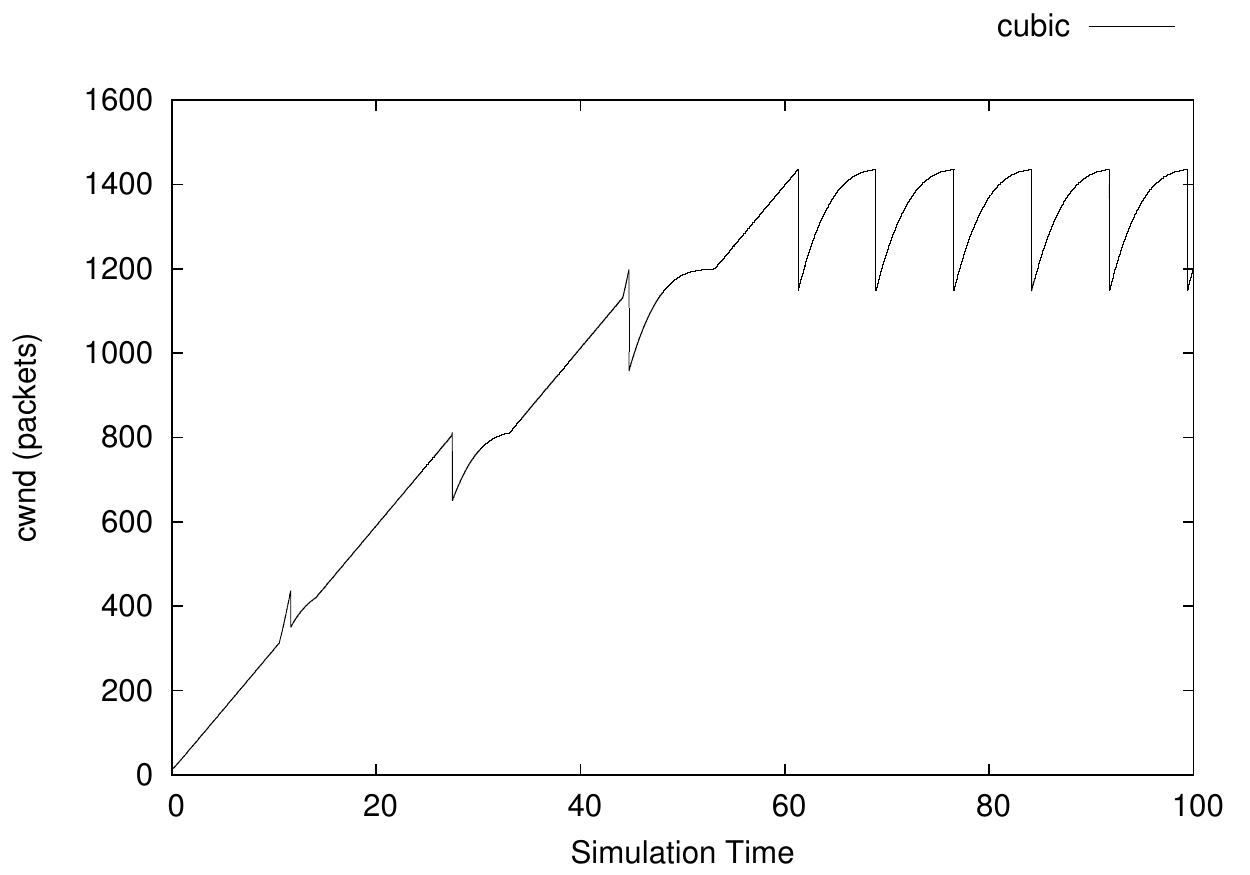}
			\label{fig:CWND:1}
		}	
%		\subfigure [Buffer Size = 10 Packets.]	{
%			\includegraphics[scale=0.4]{agile2}			
%			\includegraphics[scale=0.4]{compound2}
%			\includegraphics[scale=0.4]{cubic2}
%			\label{fig:CWND:2}
%		}	
		\subfigure [Buffer Size = 25 Packets.]	{
			\includegraphics[scale=0.39]{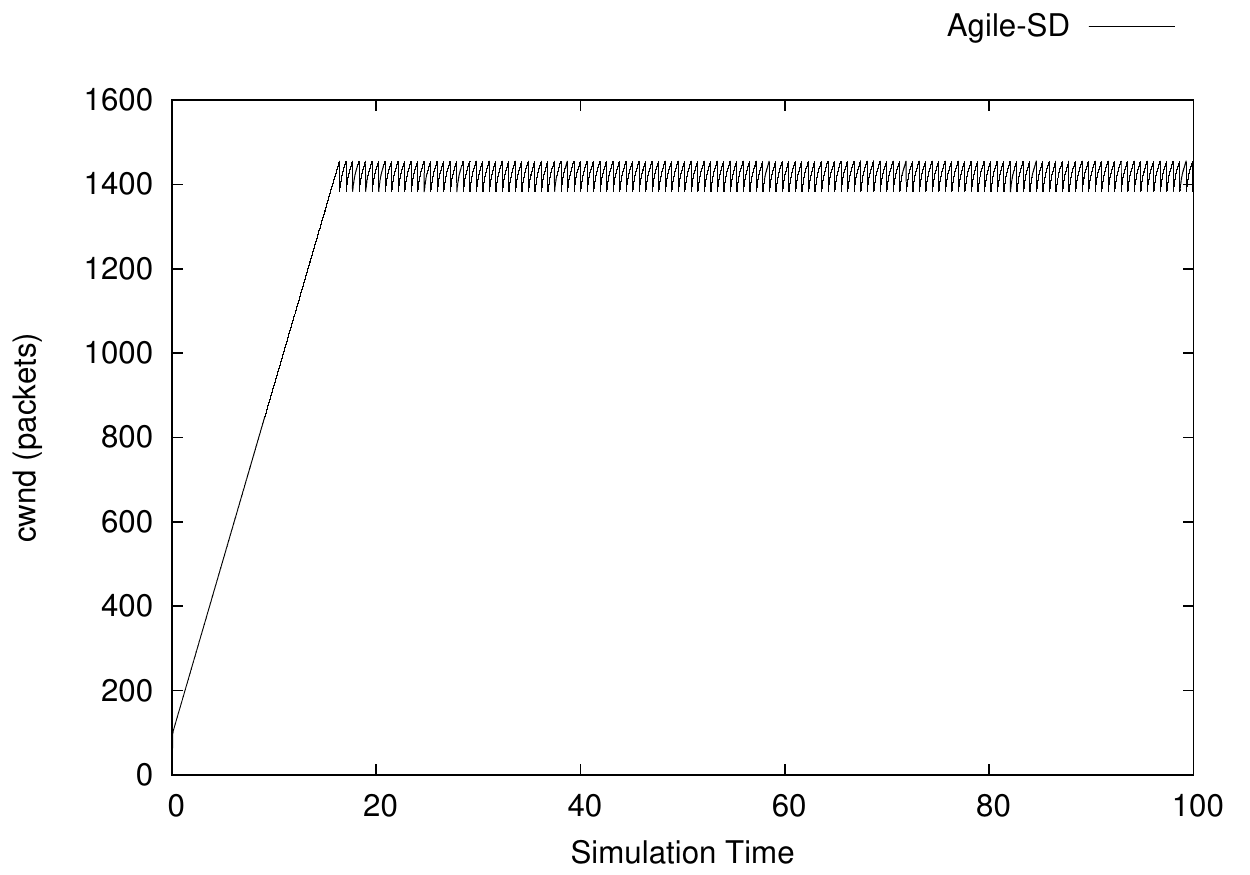}		
			\includegraphics[scale=0.39]{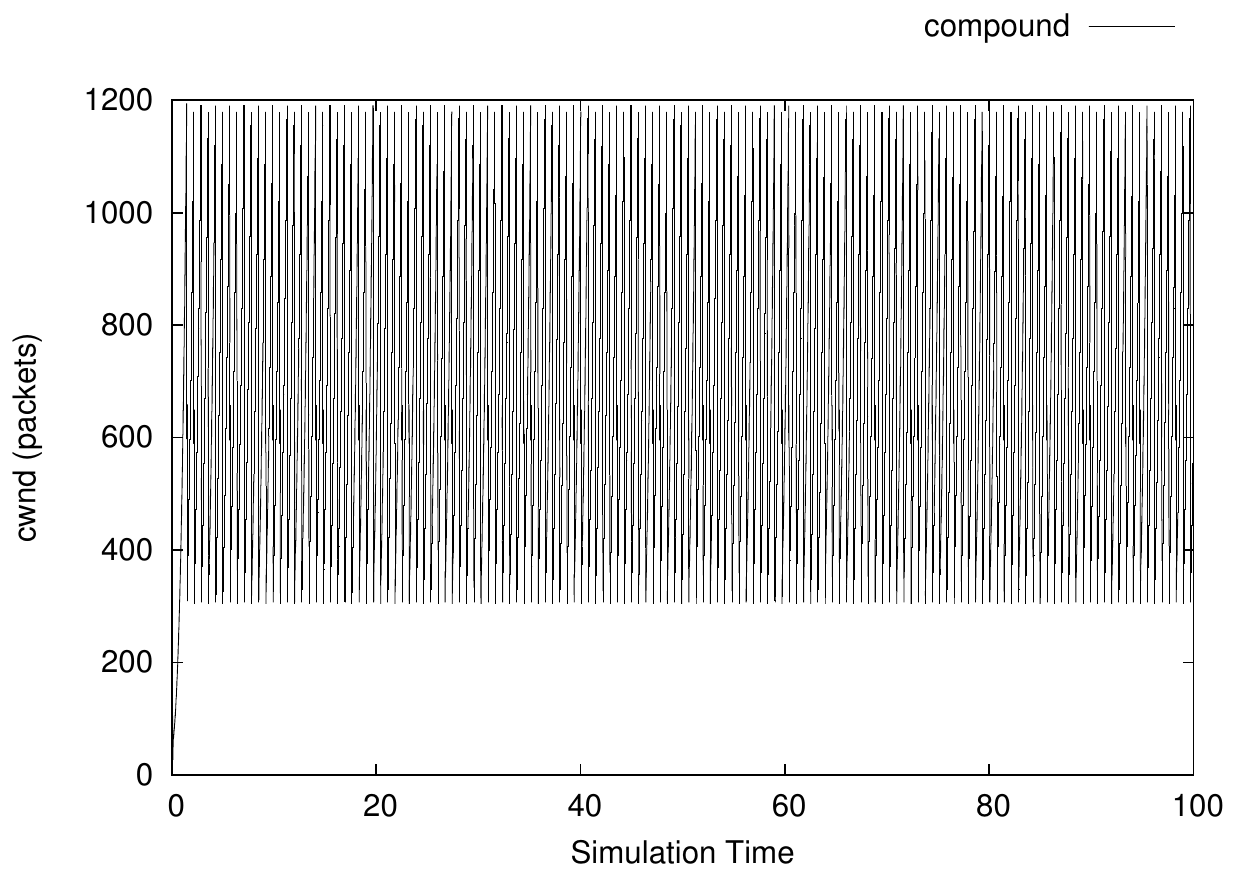}
			\includegraphics[scale=0.39]{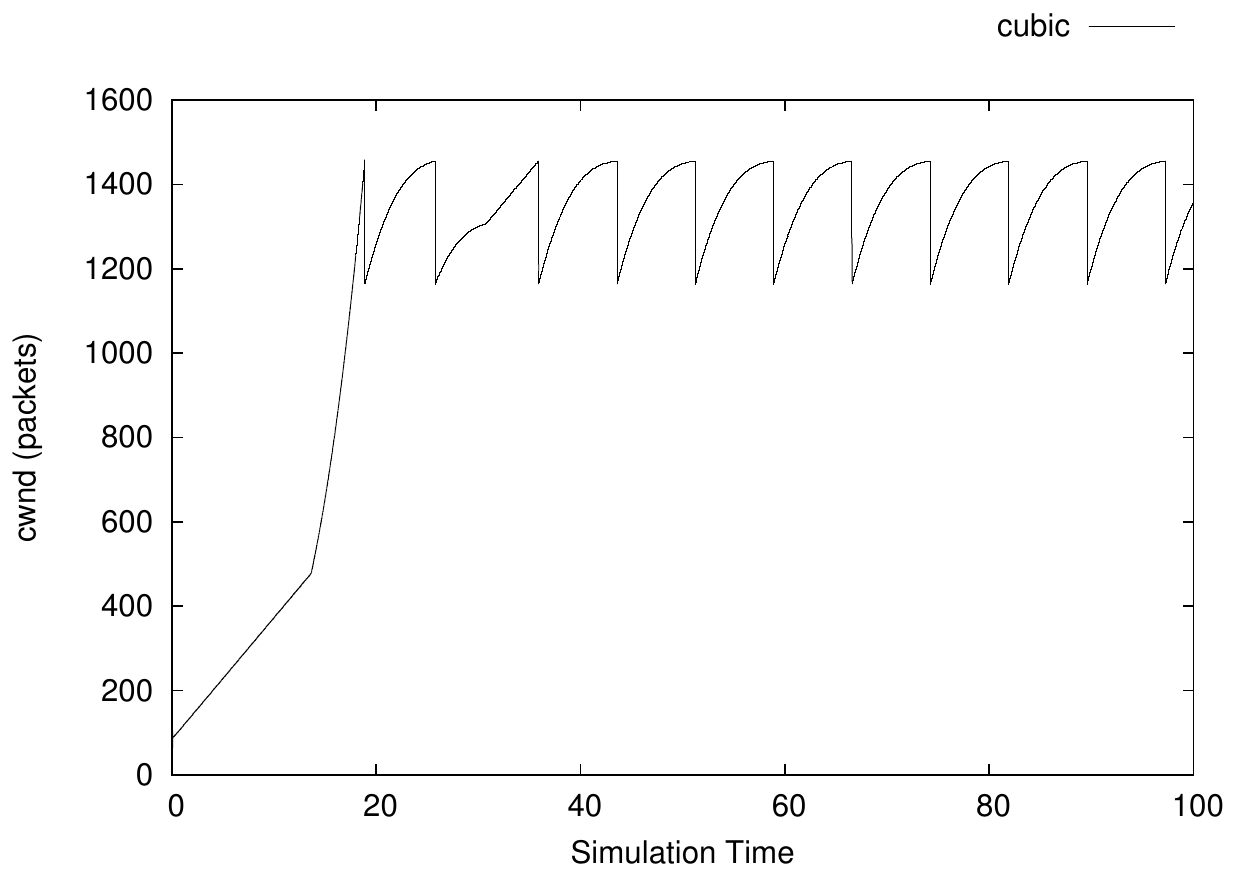}
			\label{fig:CWND:3}
		}	
%		\subfigure [Buffer Size = 50 Packets.]	{
%			\includegraphics[scale=0.4]{agile4}			
%			\includegraphics[scale=0.4]{compound4}
%			\includegraphics[scale=0.4]{cubic4}
%			\label{fig:CWND:4}
%		}	
		\subfigure [Buffer Size = 100 Packets.]	{
			\includegraphics[scale=0.39]{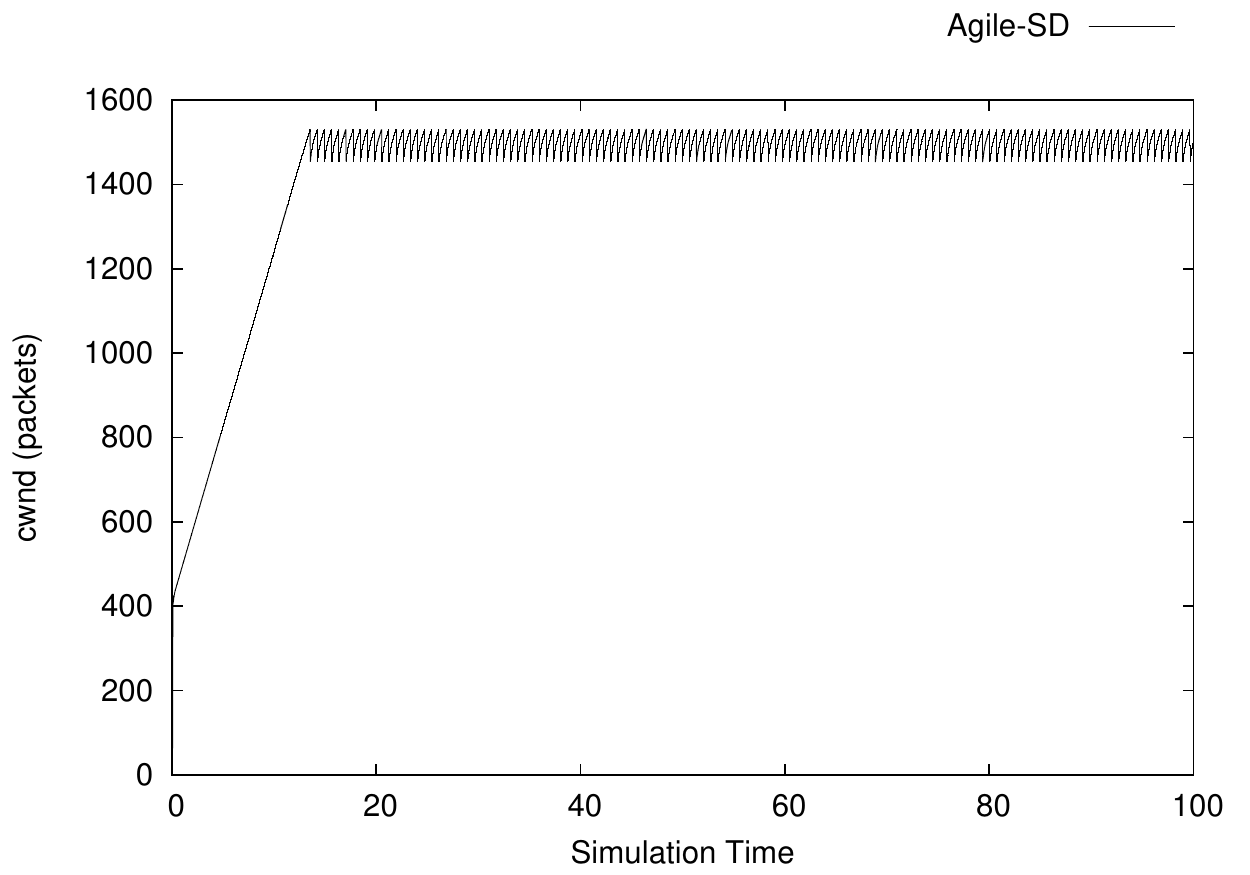}		
			\includegraphics[scale=0.39]{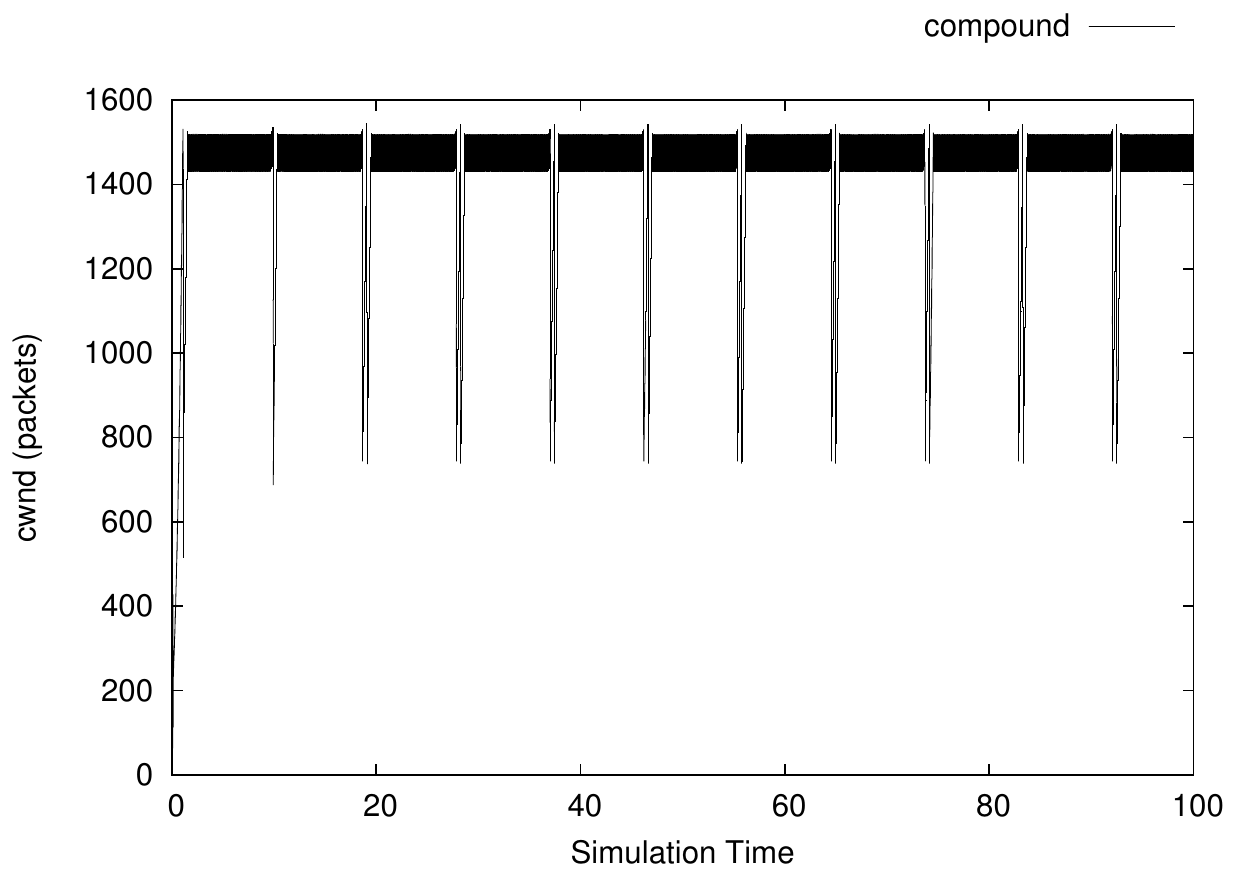}
			\includegraphics[scale=0.39]{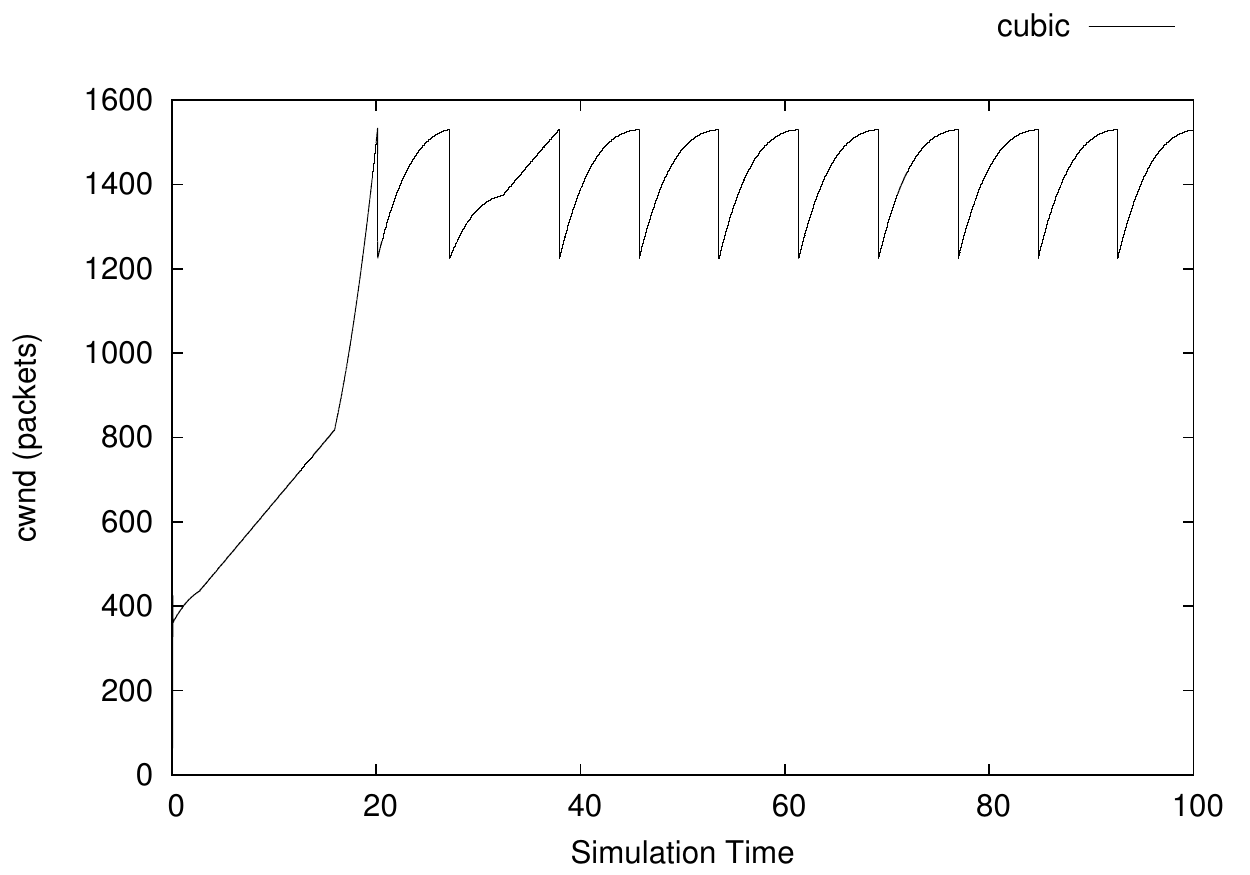}
			\label{fig:CWND:5}
		}
%	\end{center}
%\end{figure*}
%\begin{figure*} [t!]
%	\centering
%	\begin{center}
		\subfigure [Buffer Size = 250 Packets.]	{
			\includegraphics[scale=0.39]{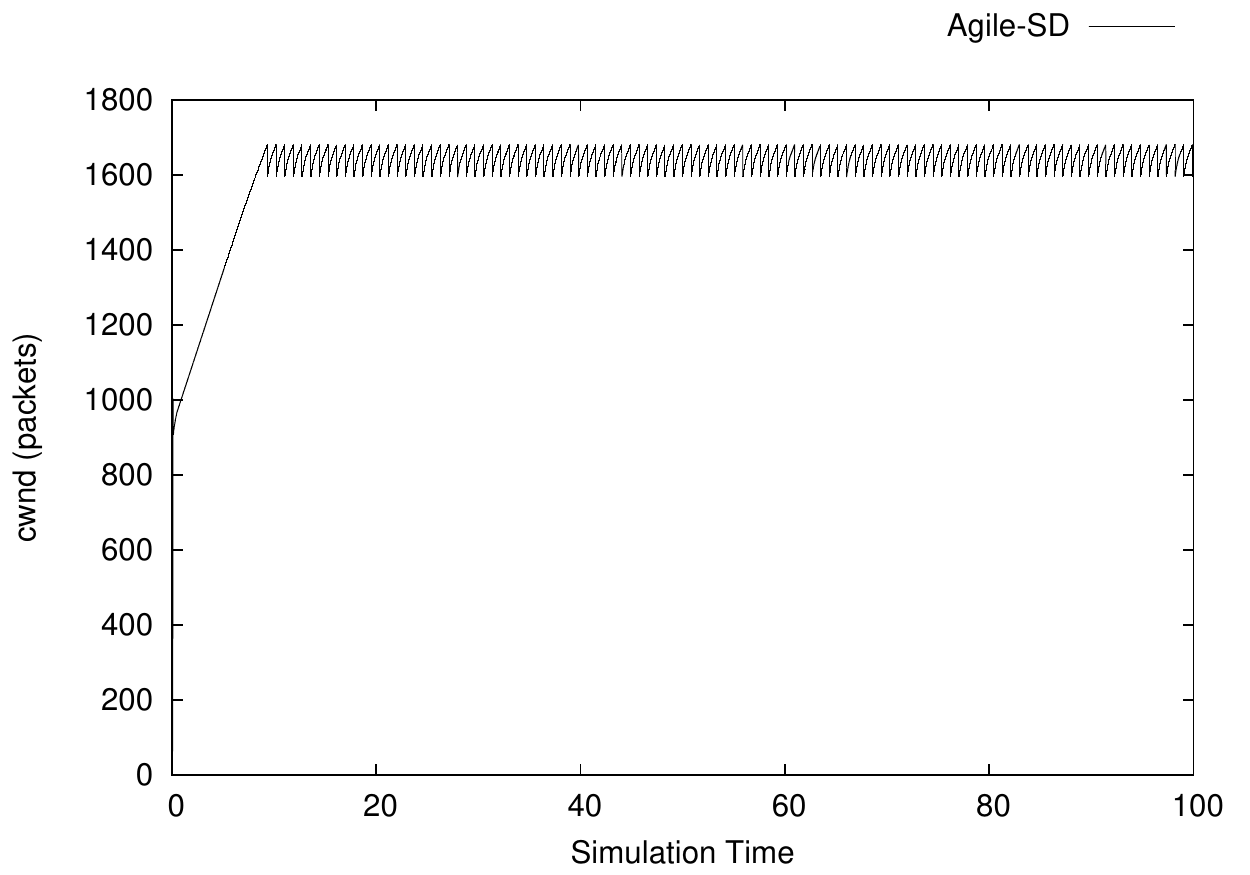}		
			\includegraphics[scale=0.39]{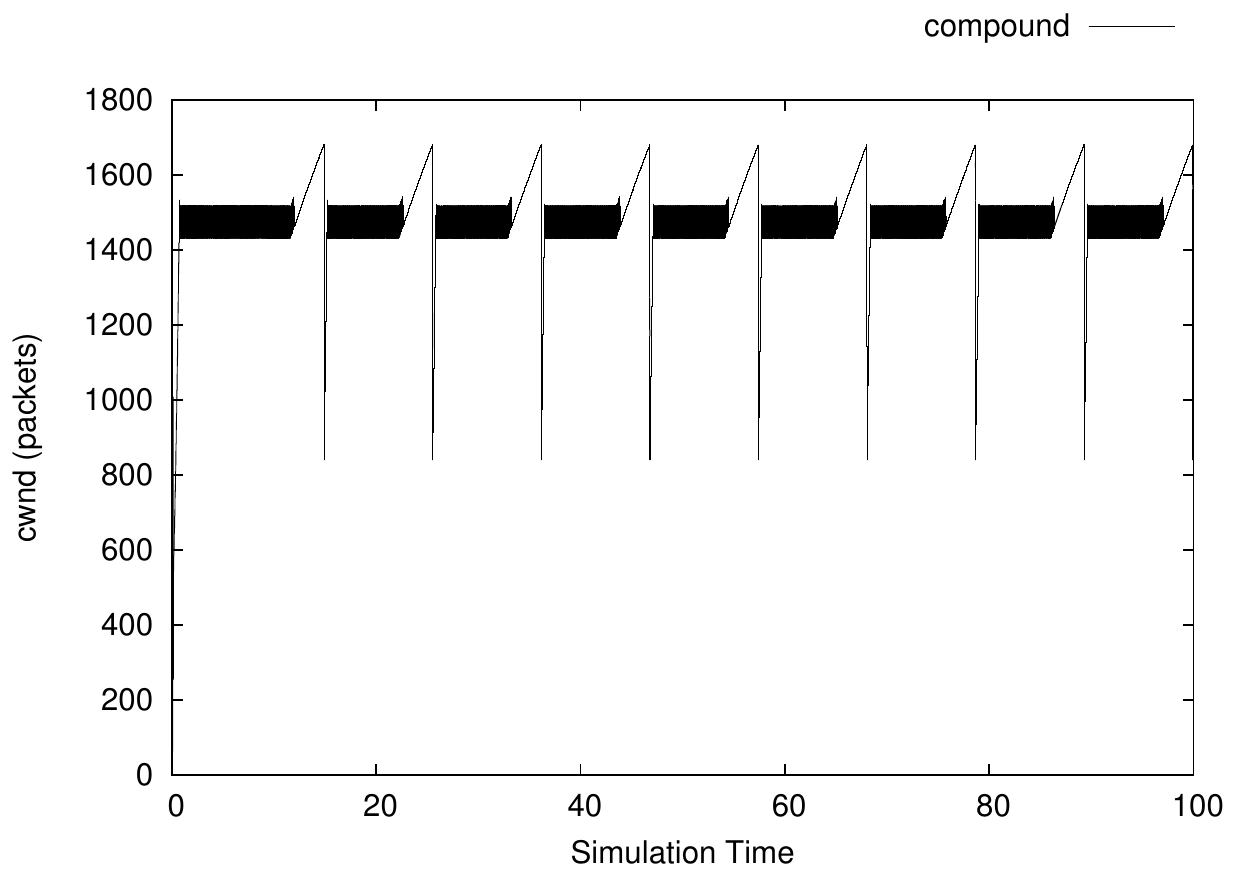}
			\includegraphics[scale=0.39]{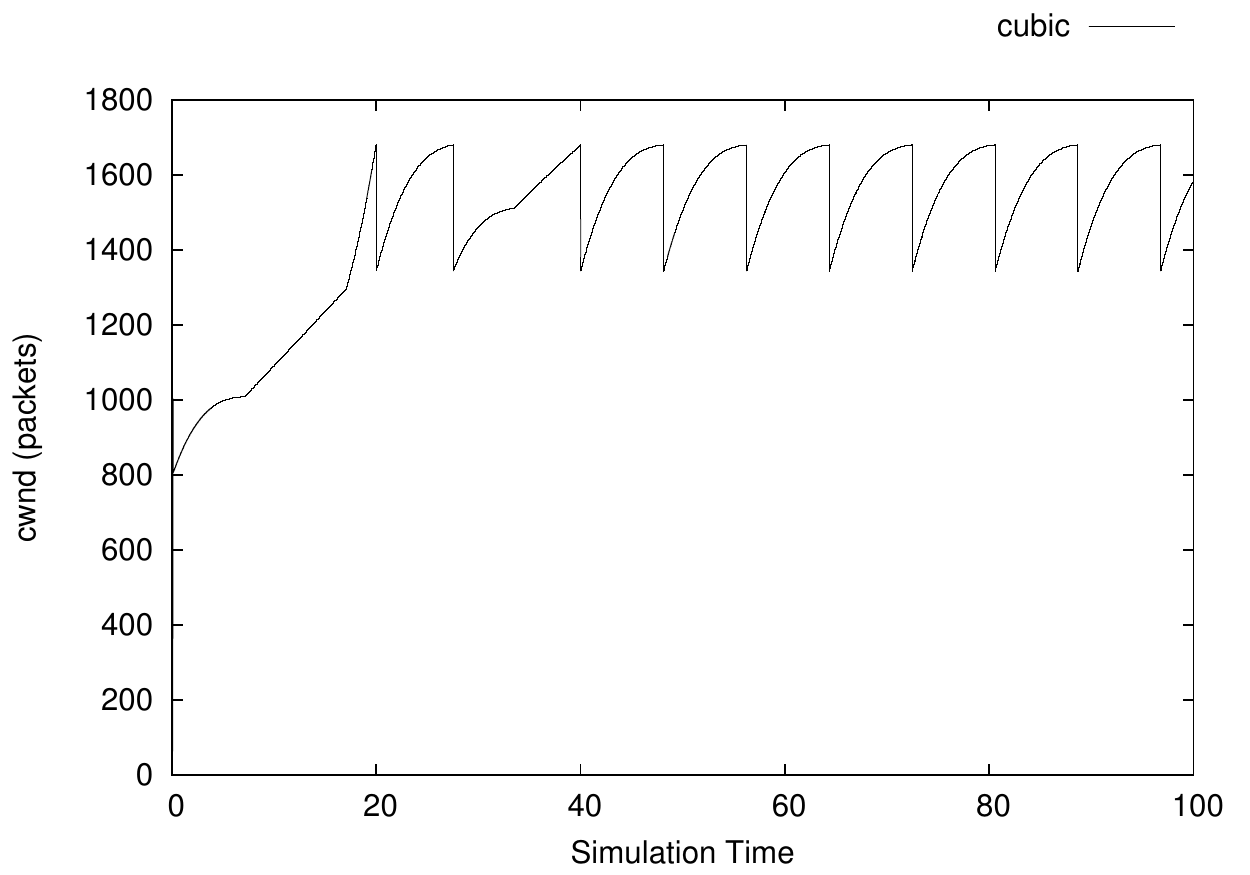}
			\label{fig:CWND:6}
		}	
		\subfigure [Buffer Size = 500 Packets.]	{
			\includegraphics[scale=0.39]{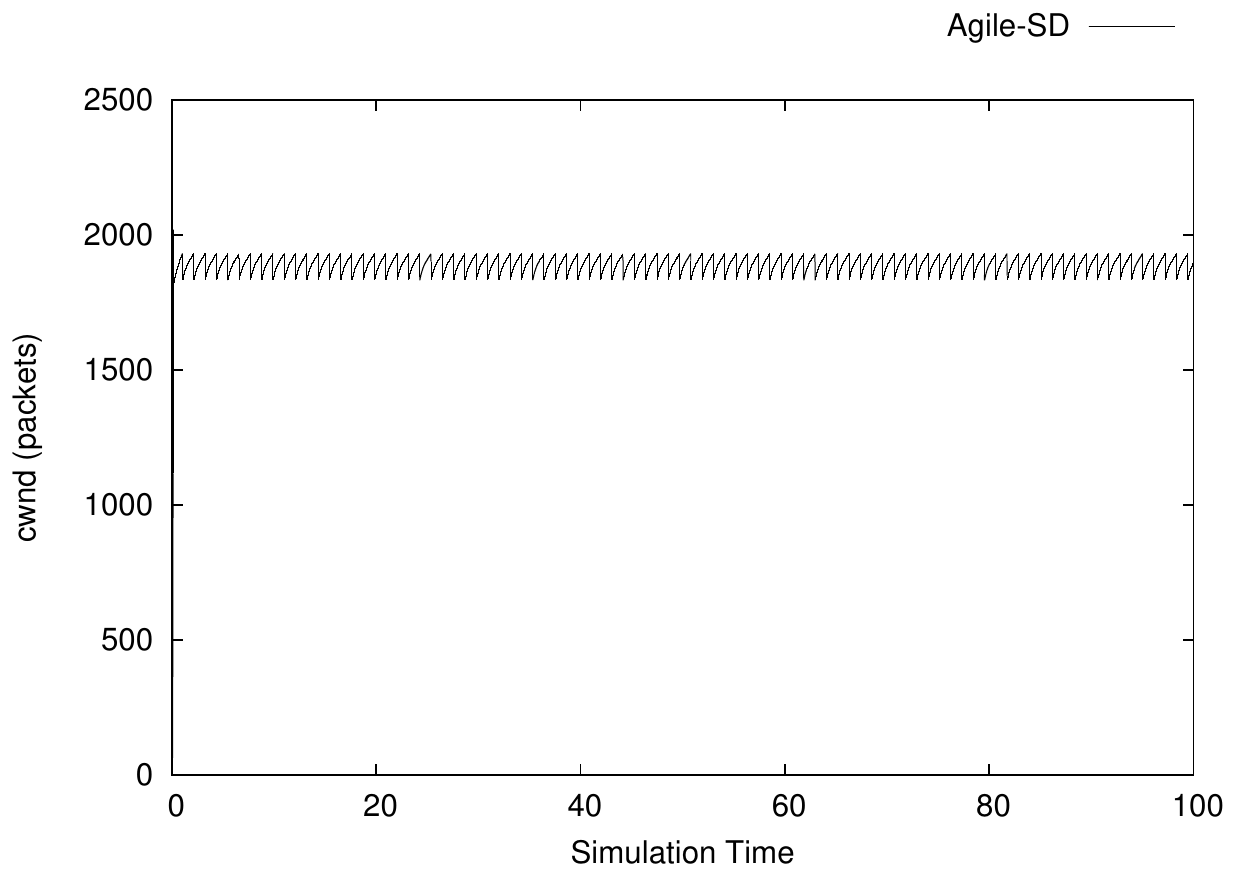}		
			\includegraphics[scale=0.39]{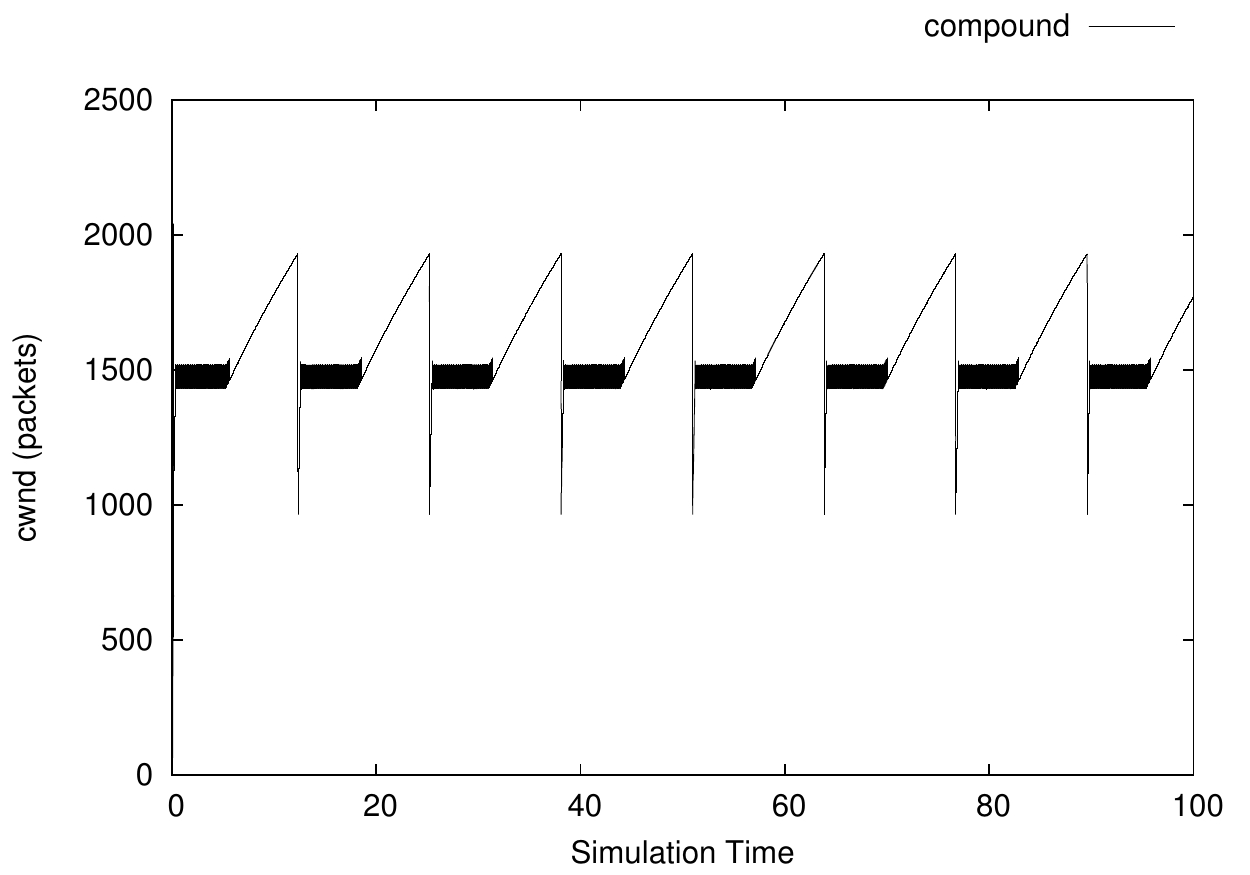}
			\includegraphics[scale=0.39]{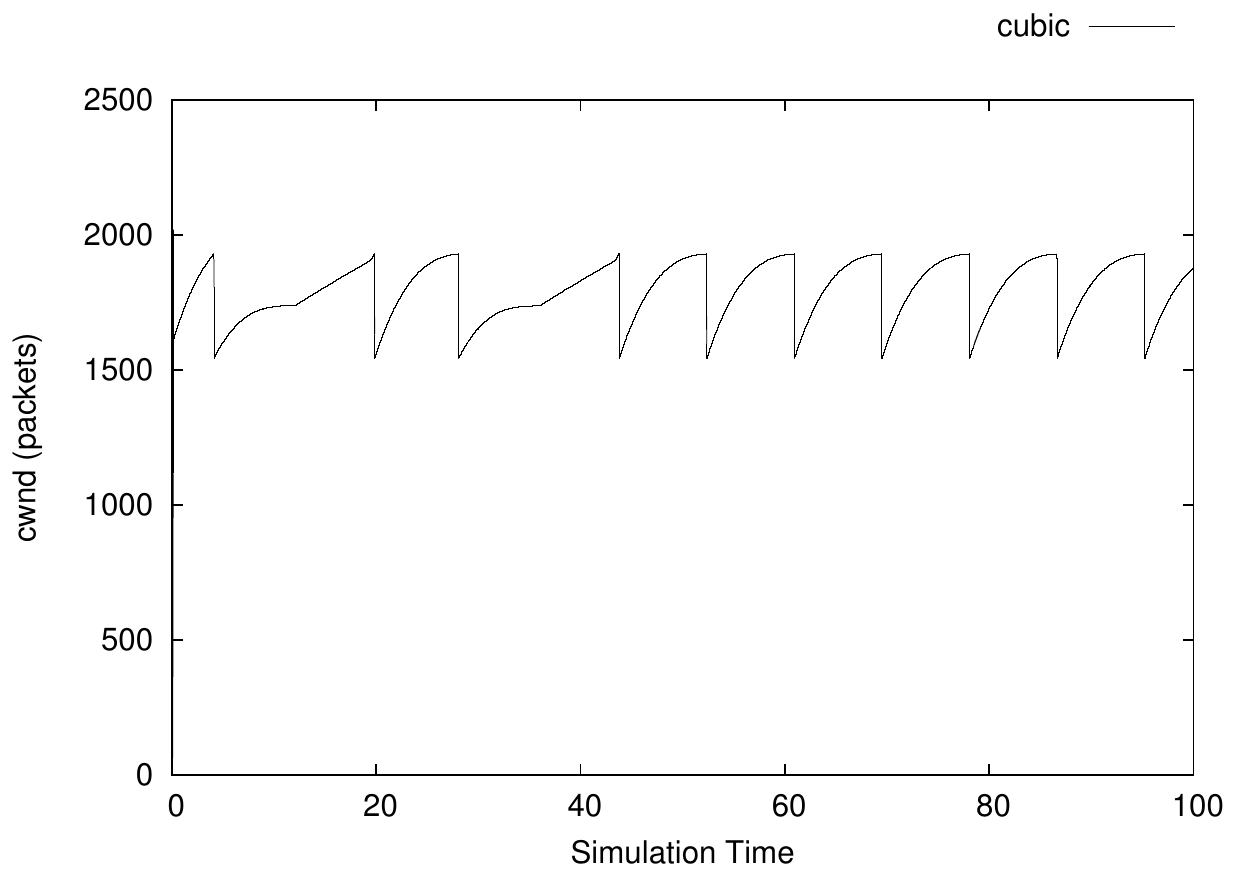}
			\label{fig:CWND:7}
		}
	\end{center}
	\caption{TCP Congestion Window Evolution.}
	\label{fig:CWND}
\end{figure*}

Fundamentally, the target of CCAs is: to maximize the throughput while minimizing the loss ratio and maintaining the fairness. Figure \ref{fig:CWND} shows the $cwnd$ evolution of the studied CCAs based on the buffer size change. Due to the mechanism of agility factor, Agile-SD expectedly shows the faster $cwnd$ growth followed by Cubic and C-TCP. This fast or slow evolution of $cwnd$ is the core of any CCA which would directly affect the other performance metrics, such as throughput, loss ratio and it may affect the fairness as well.

In Figure \ref{fig:CWND:1}, it is very clear that Agile-SD reaches the maximum $cwnd$, which is about 1500 packets, in around 17 second then starts oscillating to show very short epochs, while, Cubic reaches the maximum $cwnd$ in about 60 seconds then starts oscillating to draw very long epochs. As for C-TCP, it fails to reach the maximum $cwnd$ and touches only the edge of 110 packets then exhibits very short epochs. Indeed, the larger the $cwnd$, the higher the throughput and vice versa.

Interestingly, when the buffer size increases, the behavior of the studied CCAs relatively improves. The figures from \ref{fig:CWND:3} to \ref{fig:CWND:7} show that Agile-SD and Cubic reduce their time of reaching the maximum $cwnd$ whenever the buffer size increases. Besides, Agile-SD keeps showing short epochs while Cubic remains drawing long epochs. Unlikely, C-TCP heightens its $cwnd$ towards the maximum limit whenever the buffer size increases. In fact, the wider oscillating and/or the longer epochs is the lower bandwidth utilization and vice versa. Thus, the higher bandwidth utilization among the studied CCAs would be provided by Agile-SD followed by Cubic then C-TCP.

\subsubsection{The average throughput}
\begin{figure*}[t!]
	\begin{center}
		\subfigure[The First Scenario: $Zero$ PER.] 
		{
			\includegraphics[scale=0.39]{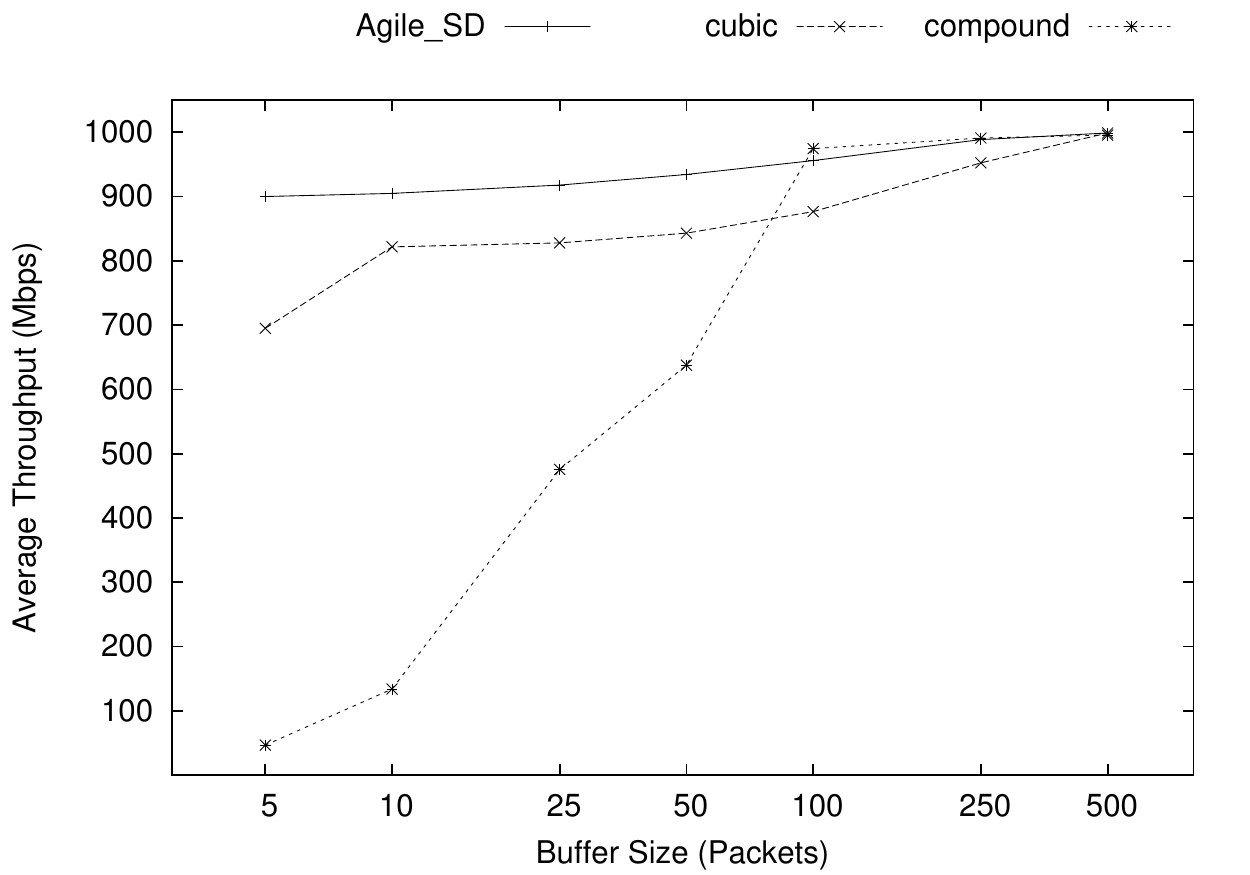}
			\label{fig:single-0per-throughput}
		}
%		\subfigure[The First Scenario: $10^{-6}$ PER.] 
%		{
%			\includegraphics[scale=0.39]{single1perthroughput}
%			\label{fig:single-1per-throughput}
%		}
		\subfigure[The First Scenario: $10^{-5}$ PER.]
		{
			\includegraphics[scale=0.39]{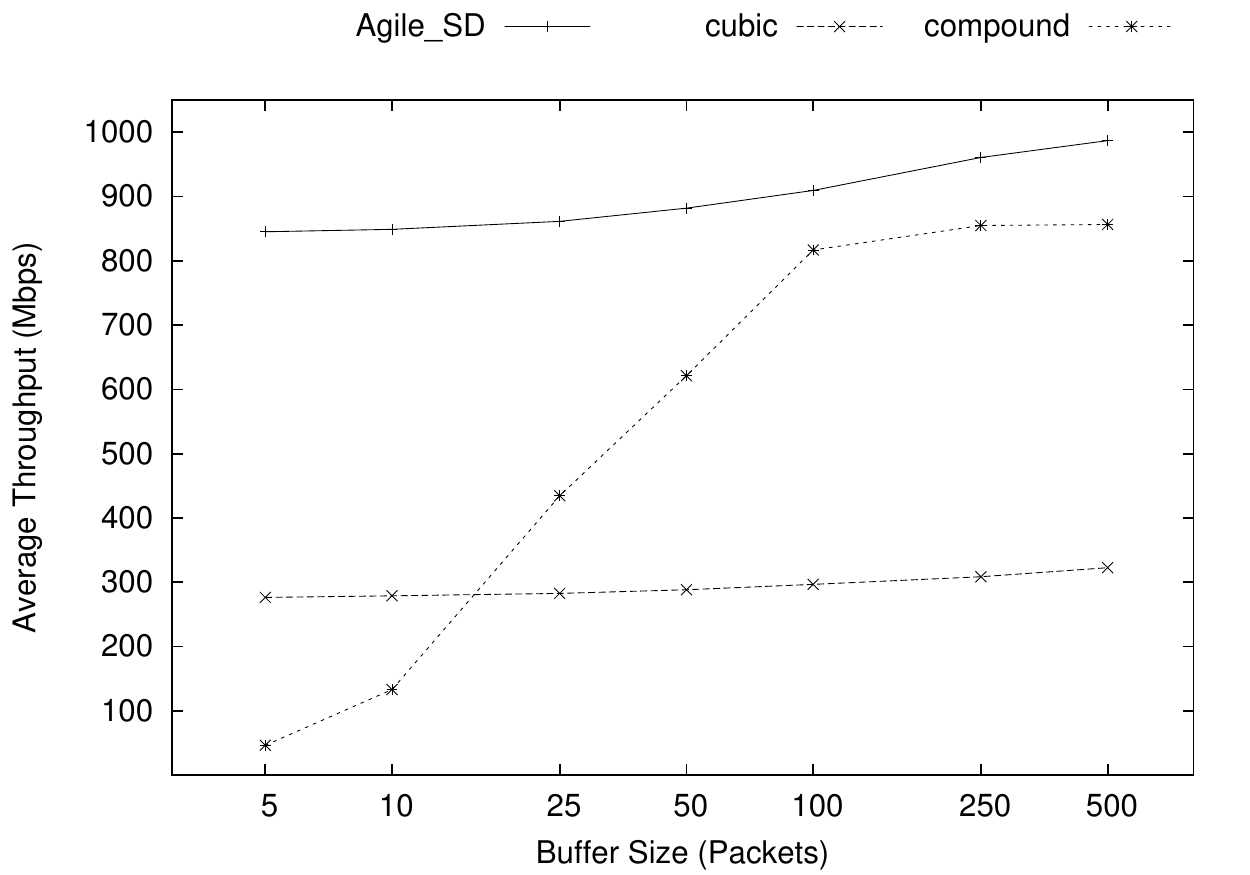}
			\label{fig:single-2per-throughput}
		}
		\subfigure[The First Scenario: $10^{-4}$ PER.]
		{
			\includegraphics[scale=0.39]{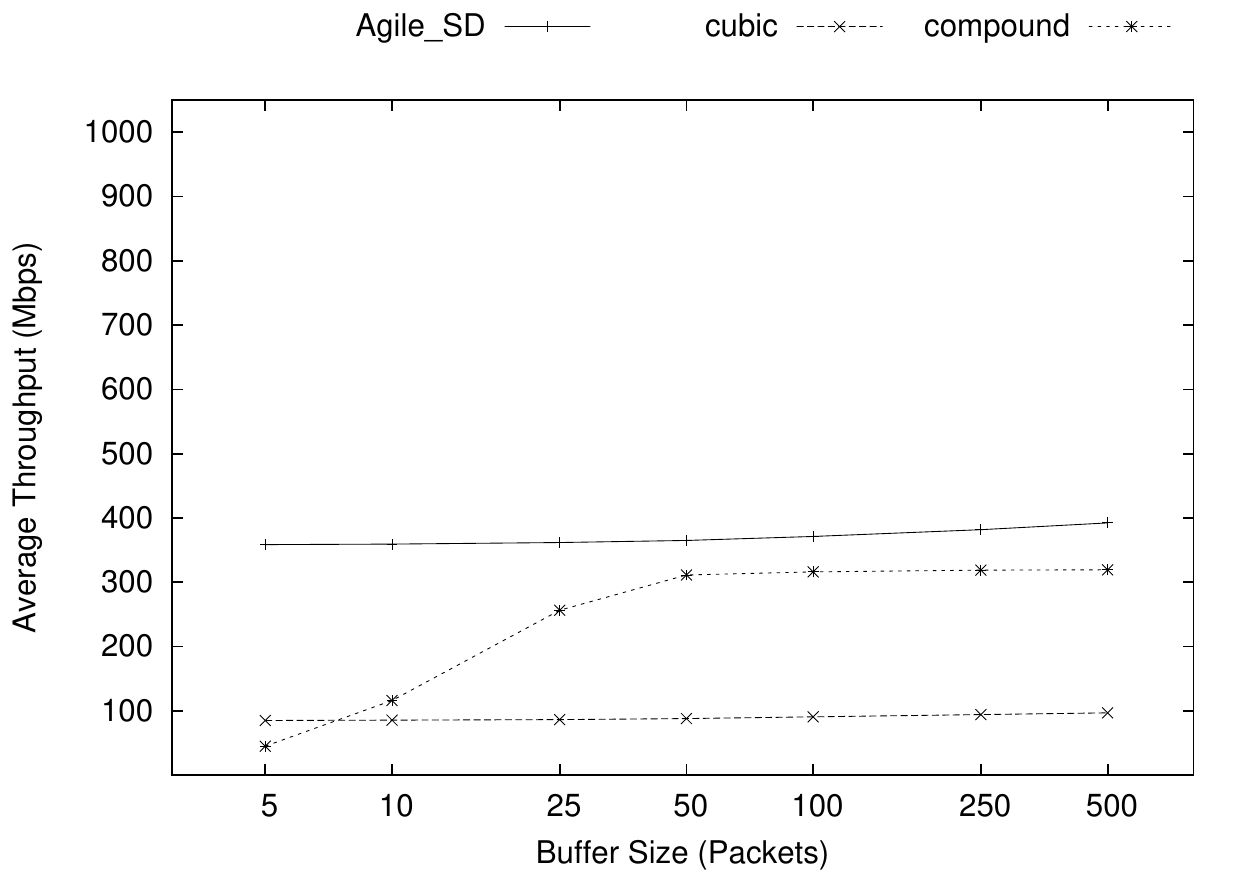}
			\label{fig:single-3per-throughput}
		}
%----------------------------------------------------------------
		\subfigure[The Second Scenario: $Zero$ PER.] 
		{
			\includegraphics[scale=0.39]{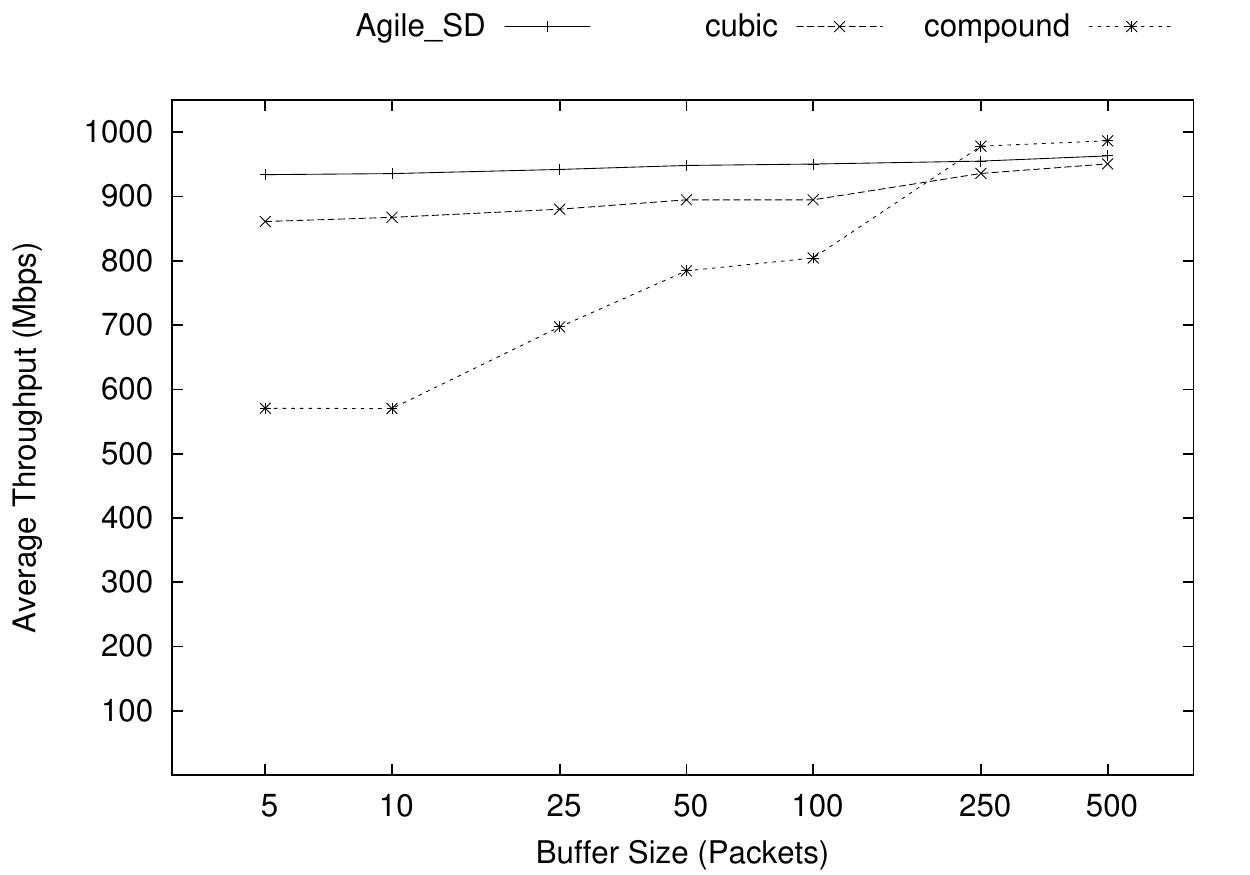}
			\label{fig:seq-0per-throughput}
		}
%		\subfigure[The Second Scenario: $10^{-6}$ PER.] 
%		{
%			\includegraphics[scale=0.39]{sperthroughput.pdf}
%			\label{fig:seq-1per-throughput}
%		}
		\subfigure[The Second Scenario: $10^{-5}$ PER.]
		{
			\includegraphics[scale=0.39]{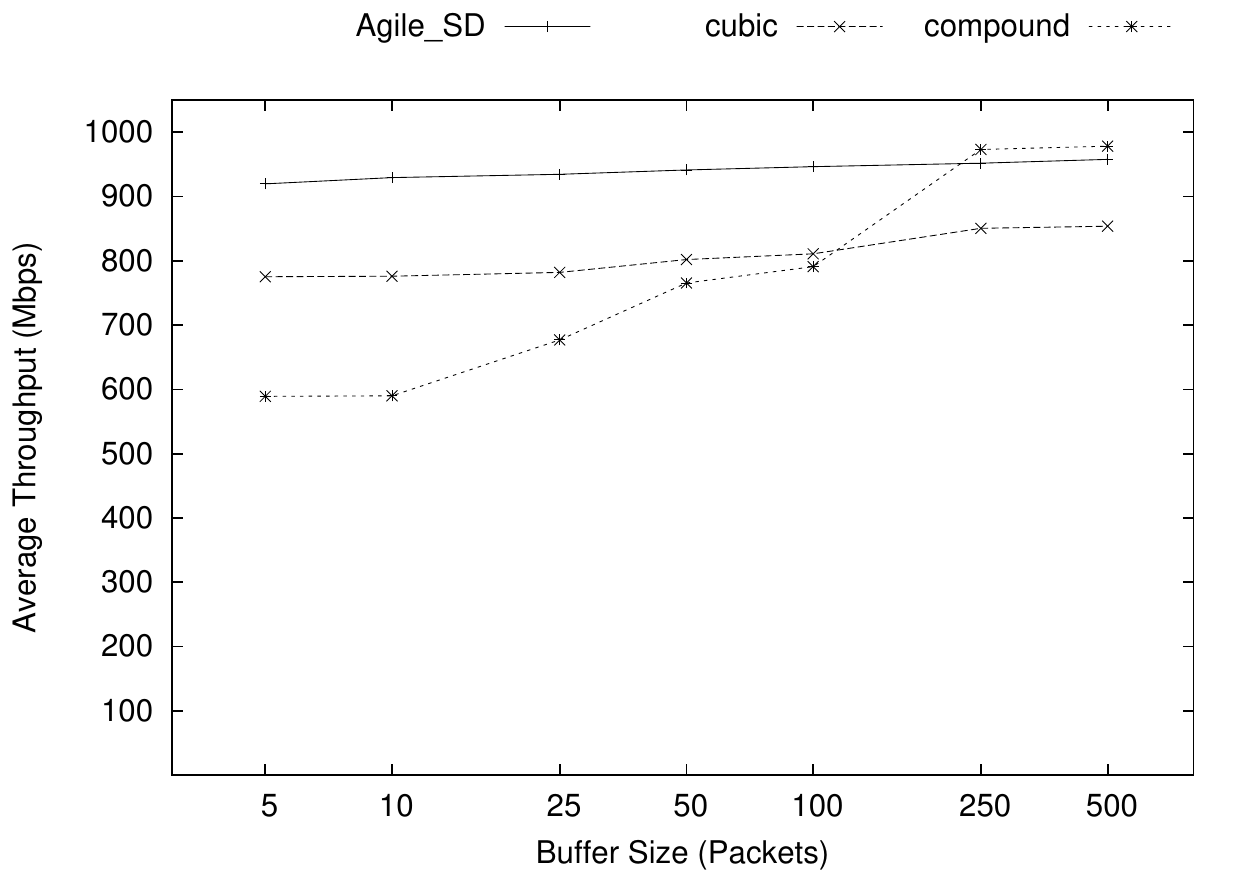}
			\label{fig:seq-2per-throughput}
		}
		\subfigure[The Second Scenario: $10^{-4}$ PER.]
		{
			\includegraphics[scale=0.39]{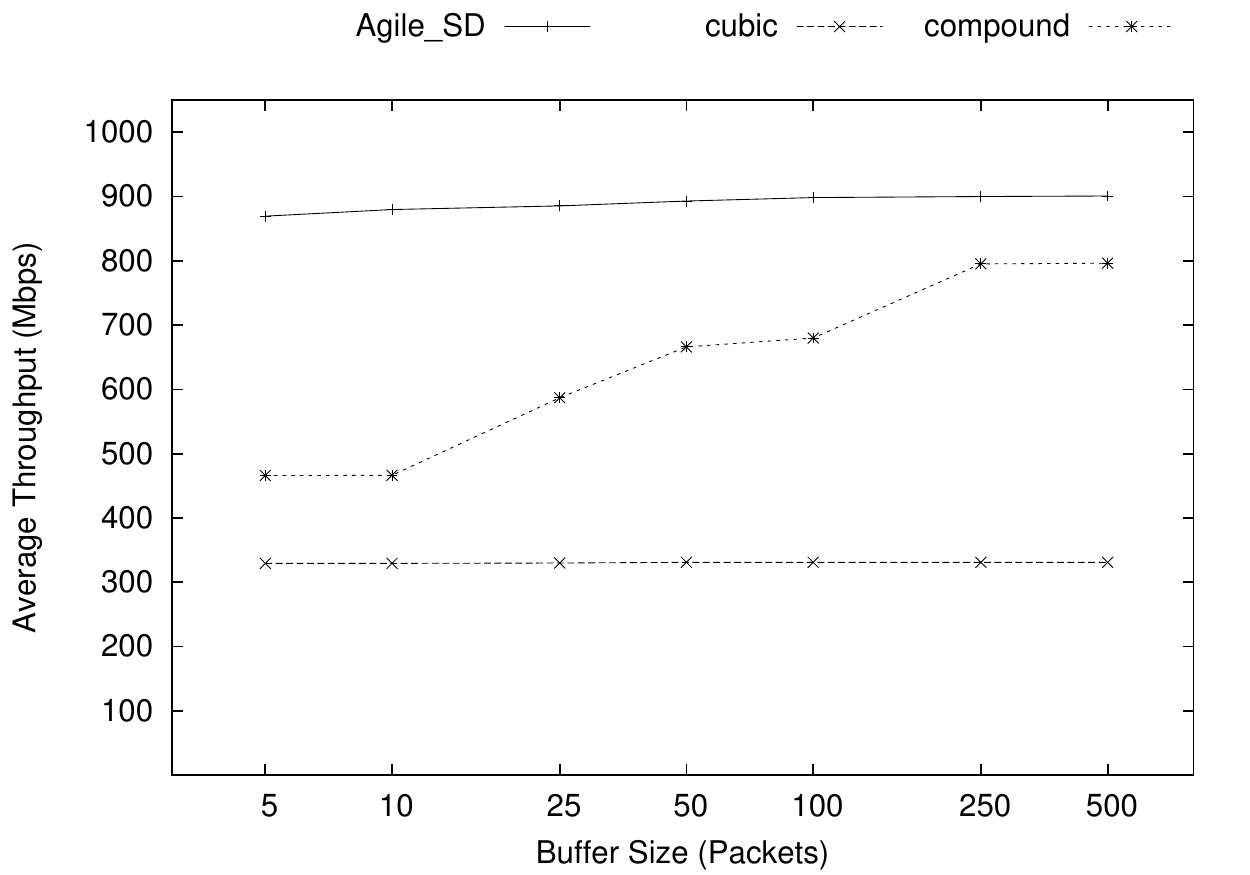}
			\label{fig:seq-3per-throughput}
		}
%----------------------------------------------------------------
		\subfigure[The Third Scenario: $Zero$ PER.] 
		{
			\includegraphics[scale=0.39]{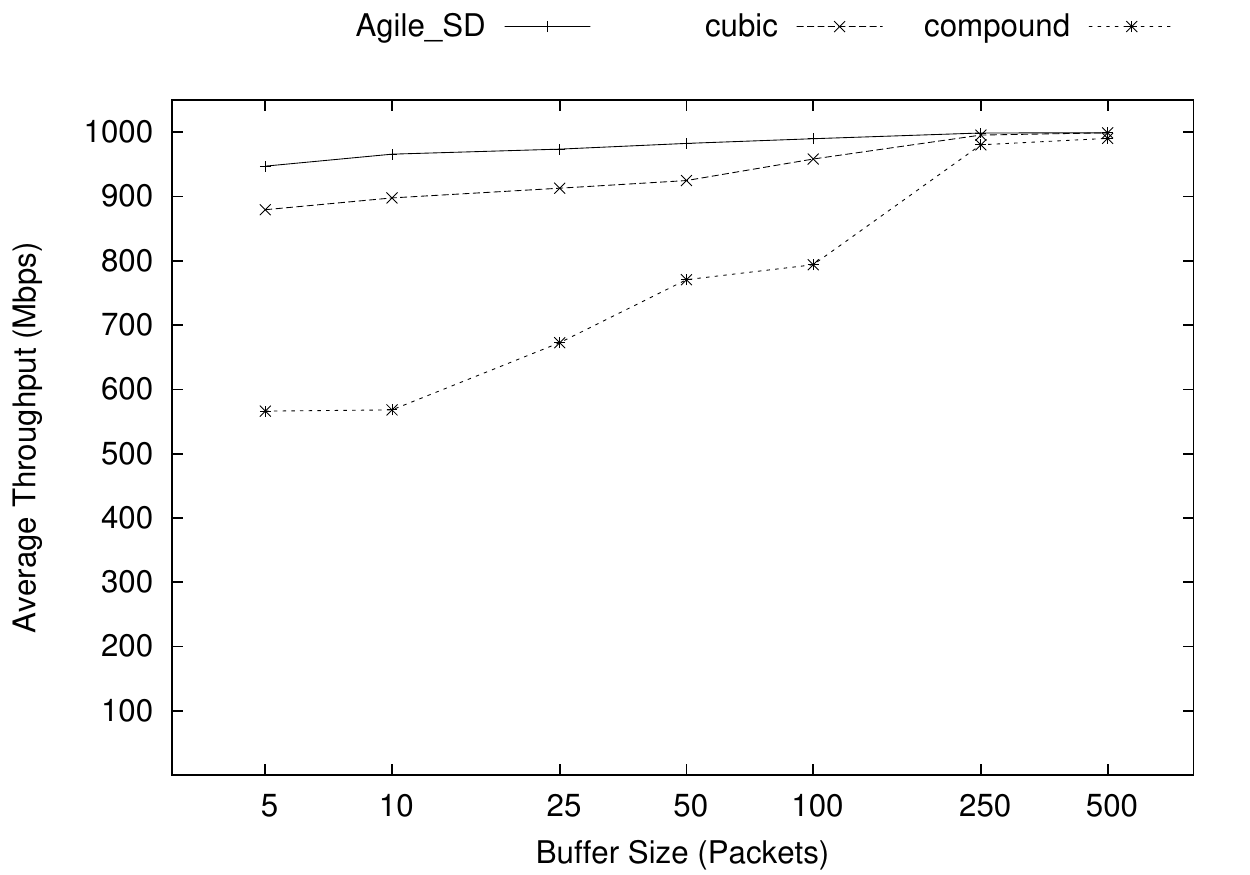}
			\label{fig:sync-0per-throughput}
		}
%		\subfigure[The Third Scenario: $10^{-6}$ PER.] 
%		{
%			\includegraphics[scale=0.39]{sync1perthroughput}
%			\label{fig:sync-1per-throughput}
%		}
		\subfigure[The Third Scenario: $10^{-5}$ PER.]
		{
			\includegraphics[scale=0.39]{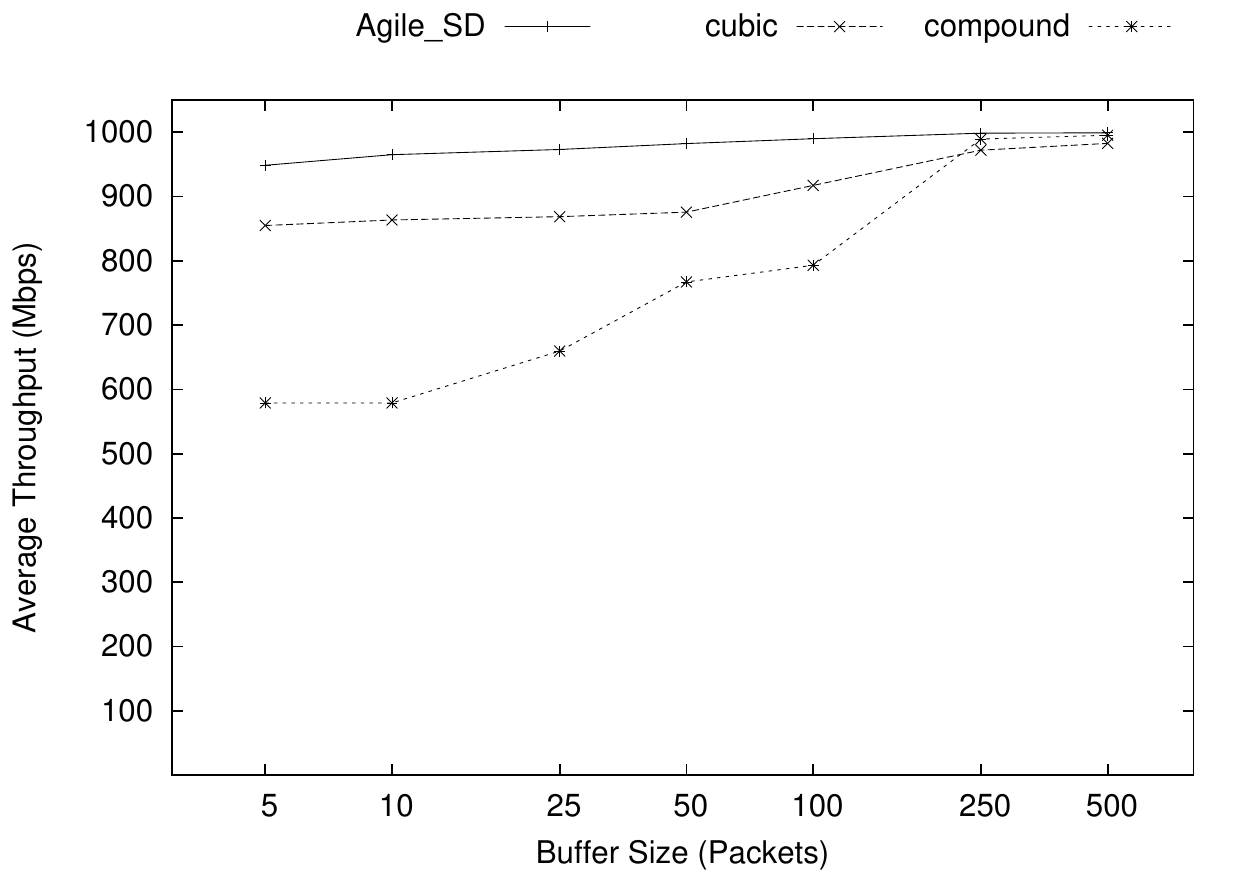}
			\label{fig:sync-2per-throughput}
		}
		\subfigure[The Third Scenario: $10^{-4}$ PER.]
		{
			\includegraphics[scale=0.39]{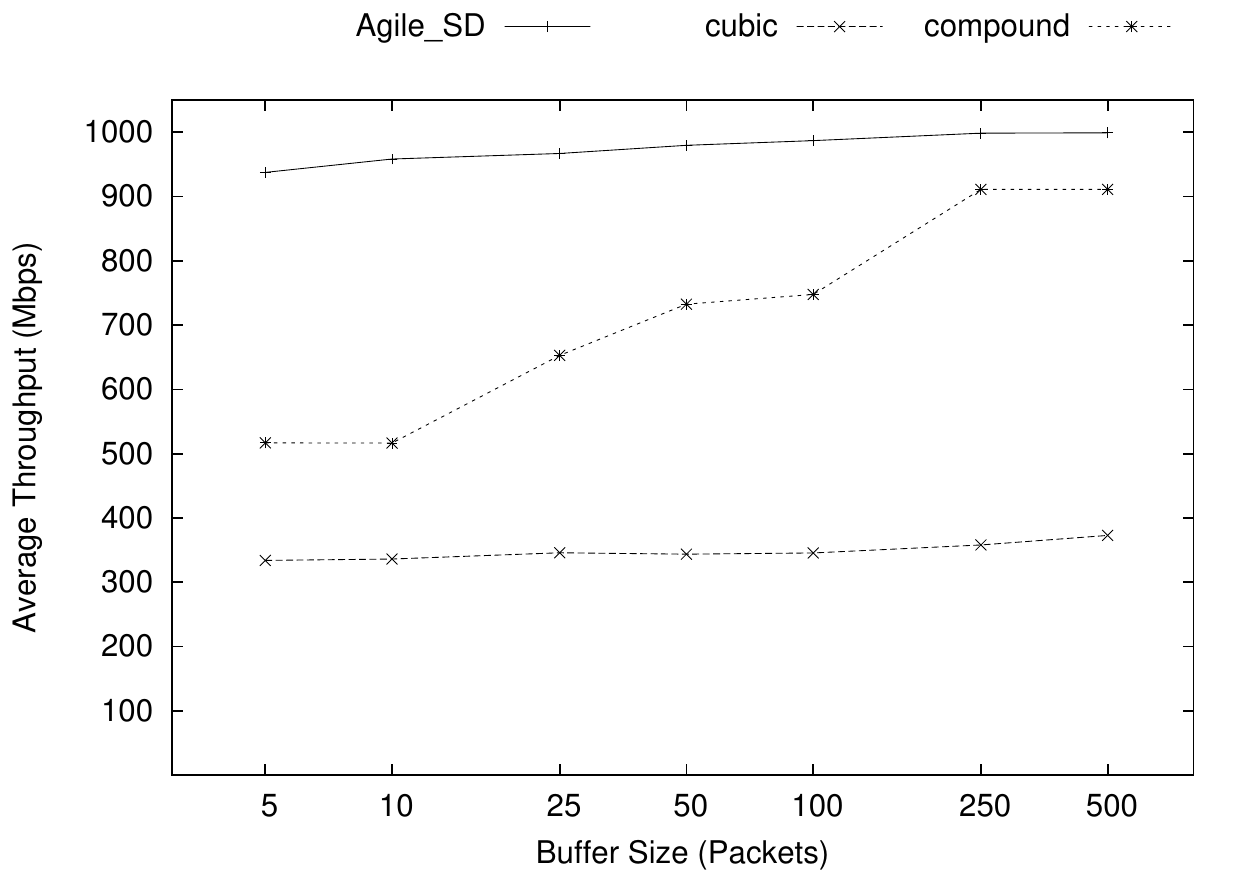}
			\label{fig:sync-3per-throughput}
		}
	\end{center}
	\caption{The Average Throughput vs. Buffer Size.}
	\label{fig:throughput}
\end{figure*}

In the first scenario, as shown in Figure \ref{fig:single-0per-throughput}, Agile-SD has overcome the other CCAs in terms of average throughput due to its fast growth of $cwnd$ resulted by the mechanism of agility factor. Moreover, Agile-SD presents lower sensitivity to PER than the others, whereas, Cubic and C-TCP are highly affected by PER and they present a poor performance when the PER is increased. However, in the cases of $10^{-4} and 10^{-5}$ PER as shown in figures \ref{fig:single-2per-throughput} and \ref{fig:single-3per-throughput}, C-TCP presents better performance than Cubic in most cases. In general, Agile-SD achieves better throughput than the others in most cases even in the lossy environments. Clearly, it improves the bandwidth utilization up to 55\% in some cases of this scenario.

For the second scenario, the figures \ref{fig:seq-0per-throughput}, \ref{fig:seq-2per-throughput} and \ref{fig:seq-3per-throughput} show that Agile-SD has overcome the other CCAs, in term of average throughput, at most cases even when the buffer size is small and the PER is high. Furthermore, it improves the bandwidth utilization from 10\% to 40\% in the cases of 5 packets buffer size. While in the third scenario, Agile-SD has outperformed the other CCAs in all cases especially when a \emph{near-zero} buffer is applied as shown in figures \ref{fig:sync-0per-throughput}, \ref{fig:sync-2per-throughput} and \ref{fig:sync-3per-throughput}. Moreover, Agile-SD significantly improves the bandwidth utilization even when the PER is high. Thus, it provides up to 40\% of improvement in some cases.

\subsubsection{The loss ratio}
\begin{figure}[h!]
\centering
\includegraphics[width=1\linewidth]{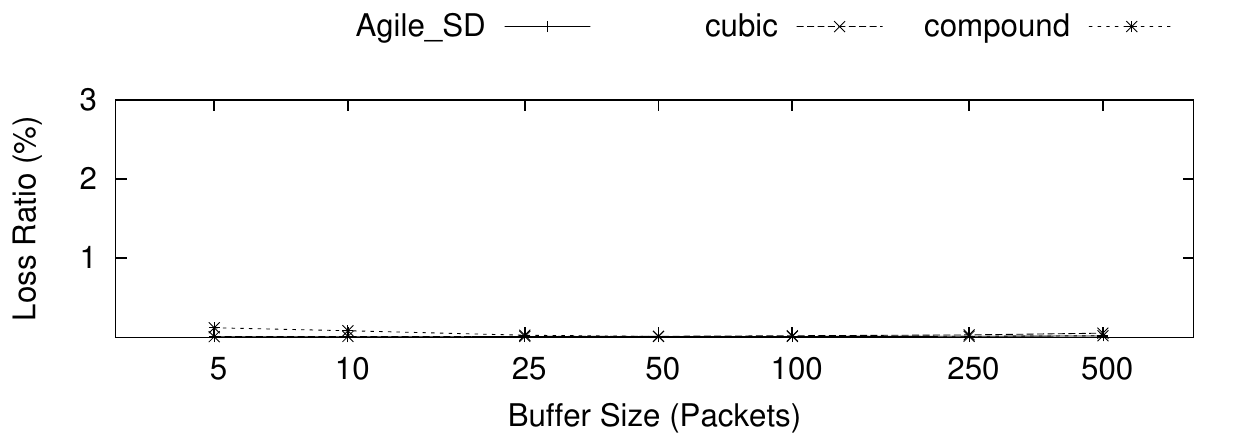}
\caption{1st Scenario ($10^{-4}$ PER): Loss Ratio vs. Buffer Size.}
\label{fig:single3perloss}
\end{figure}

As regarding to all scenarios, the studied CCAs have presented a loss ratio lower than 0.5\% which is considered as a negligible loss ratio. Thus, Figure \ref{fig:single3perloss} has been selected here as a sample of the loss ratio results of all scenarios to save the space of this paper and because the rest of the figures have no much difference.

\subsubsection{The fairness}
\begin{figure}[h!]
\centering
\includegraphics[width=\linewidth]{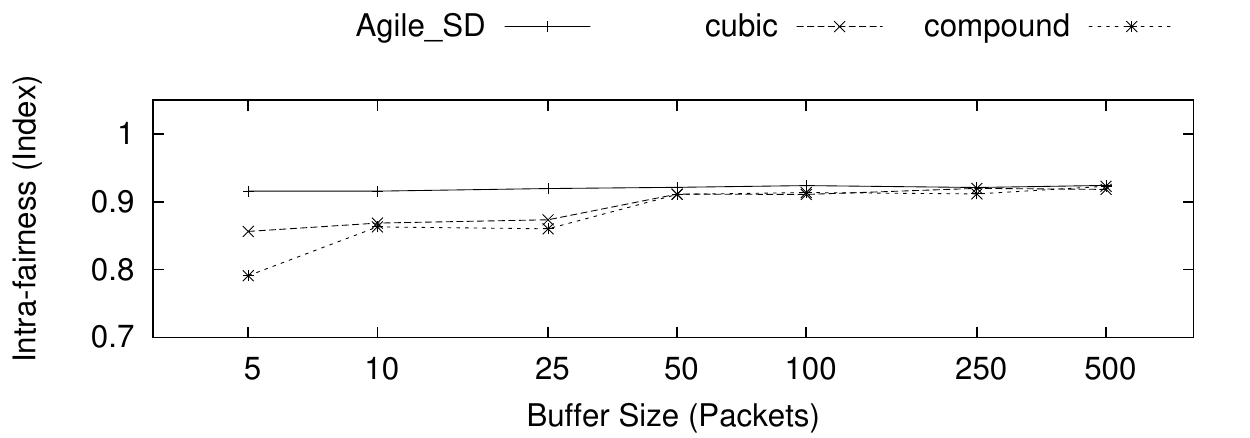}
\caption{2nd Scenario ($Zero$ PER): Intra-Fairness vs. Buffer Size.}
\label{fig:seq0perintra}
\end{figure}

\begin{figure}[h!]
\centering
\includegraphics[width=\linewidth]{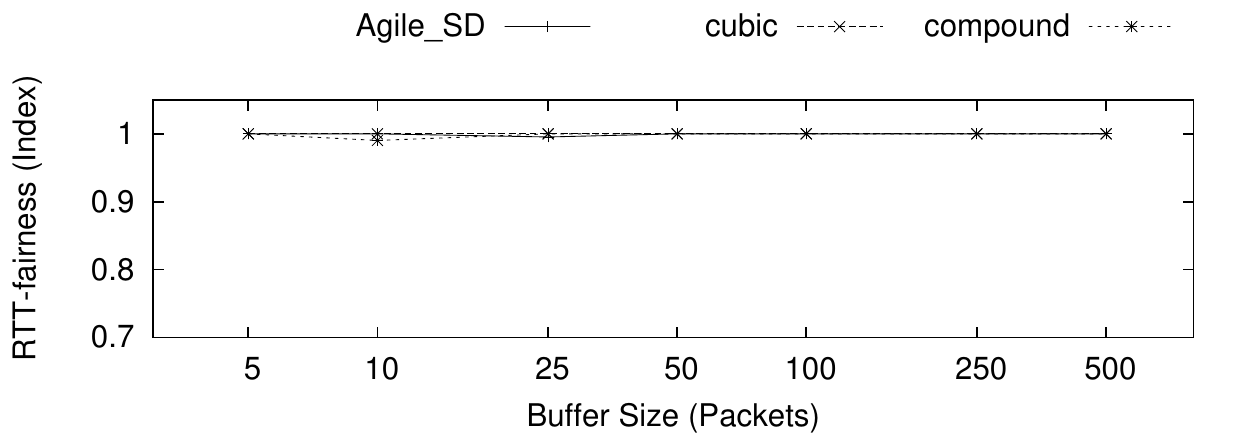}
\caption{3rd Scenario ($Zero$ PER): RTT-Fairness vs. Buffer Size.}
\label{fig:sync0perrtt}
\end{figure}

As for intra-fairness and RTT-fairness, all the studied CCAs are interchangeably close to each other. However, in some cases Agile-SD seems more fair, especially when the applied buffer size is small. Since the difference among the graphs of the results is very trivial, the figures \ref{fig:seq0perintra} and \ref{fig:sync0perrtt} have been chosen as samples of intra-fairness and RTT-fairness, respectively.

\begin{figure}[h!]
\centering
\includegraphics[angle=-90,width=0.9\linewidth]{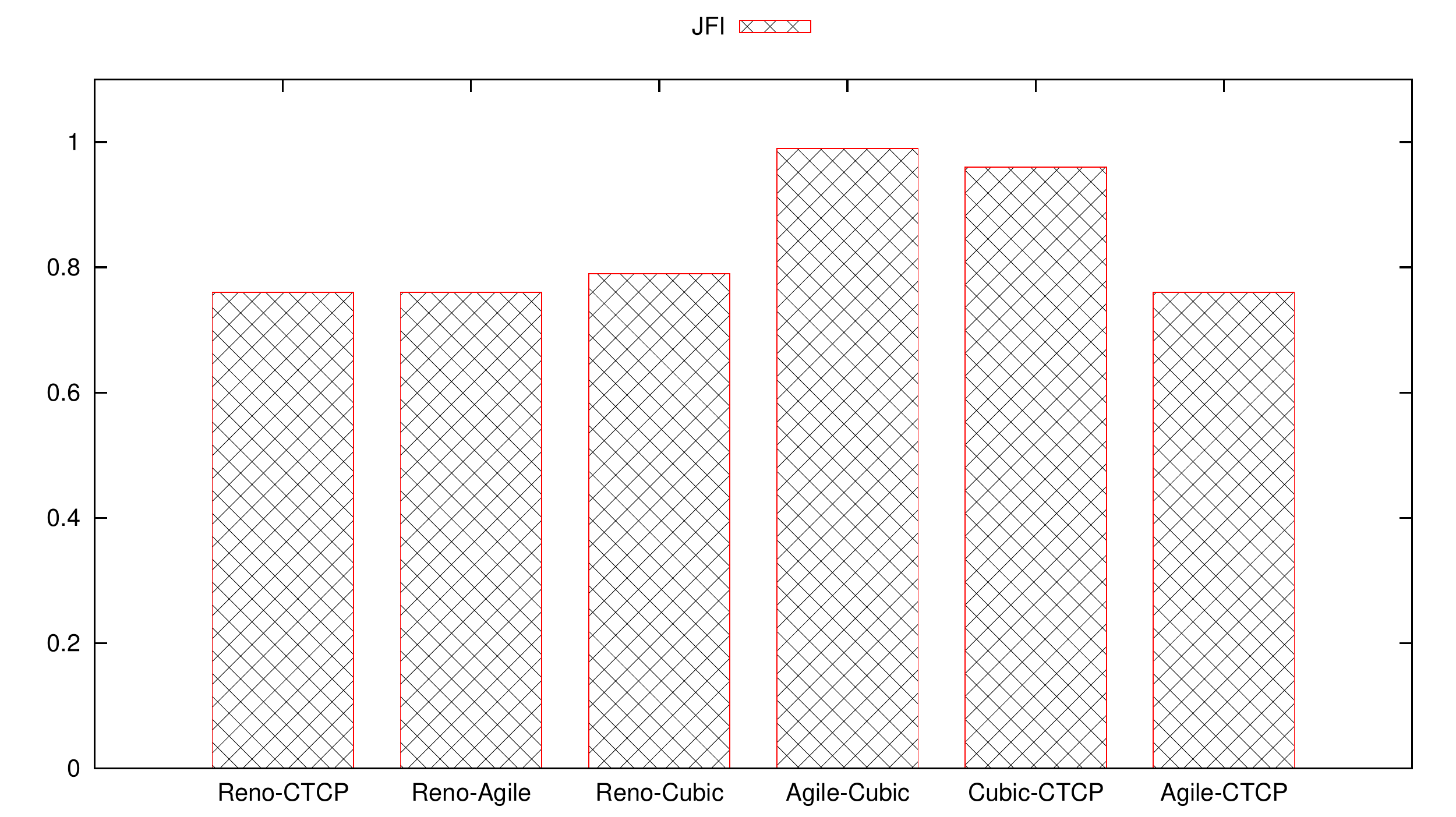}
\caption{The inter-fairness among the studied CCAs.}
\label{fig:interfairness}
\end{figure}

Moreover, a separated experiment has been carried out to evaluate the inter-fairness among the studied CCAs and the standard NewReno, using the same topology as shown in Figure \ref{fig:topology}, where the result is shown in Figure \ref{fig:interfairness}. For inter-fairness to NewReno, Agile-SD and C-TCP achieve around 0.76 while Cubic achieves about 0.79 inter-fairness index. As for the inter-fairness to Cubic, Agile-SD scores the highest index which is almost 0.99 and C-TCP scores around 0.96 while NewReno achieves only 0.79 inter-fairness index. As for the inter-fairness to C-TCP, Cubic scores the highest index which is around 0.96 while Agile-SD and NewReno achieve about 0.76 inter-fairness index.

\section{Conclusion}
\label{Conc}

In this paper, a new CCA, namely Agile-SD, has been proposed and evaluated. The main contribution of the proposed CCA is to implement the mechanism of agility factor. The need of the proposed CCA has been arisen by the inability of the existing high-speed CCAs in achieving a full bandwidth utilization over high-speed networks, especially when a small buffer regime is applied. Further, a new CCA module has been implemented and plugged into the Linux kernel version 3.19.0. As well as, this module has been plugged into the Network Simulator NS-2 version 2.35, as a Linux TCP, in order to evaluate it by comparing its performance to the other CCAs. 

Subsequently, intensive simulation experiments have been conducted to evaluate the proposed CCA by comparing its performance to C-TCP and Cubic, which are the current default TCP algorithms of MS Windows and Linux, respectively. The results show that the proposed algorithm achieves higher bandwidth utilization than the existing CCAs while maintaining fairness. Due to the use of agility factor, Agile-SD shows lower sensitivity to the changes of buffer size and PER.

Importantly, Agile-SD presents higher performance than the compared CCAs and it provides a significant improvement which is: up to 55\% in the case of single flow, up to 40\% in the case of sequentially established/terminated multi-flows and up to 40\% in the case of synchronously established/terminated multi-flows.

More importantly, the second scenario presents the real case of network, in which all TCP flows are not established or terminated synchronously. In this scenario, Agile-SD has achieved up to 95\% bandwidth utilization while the others did not exceed it in the case of large buffer. As for the case of small buffer, Agile-SD achieves around 92\% bandwidth utilization while the other TCP variants achieve from 32\% to 85\% bandwidth utilization.

Eventually, Agile-SD is a sender-side TCP module which does not change anything at receiver-side. It uses the standard slow start and provides a new congestion avoidance algorithm featured by the mechanism of agility factor. Currently, we have already implemented Agile-SD into the latest Linux kernel 3.19.0 and a real dumbbell topology has been built using Dummynet over PC-BSD version 10 to evaluate the proposed CCA based on real test-bed in the nearest future. Also, there is a strong intention to evaluate Agile-SD with SACK and/or FACK features to show their impacts on the throughput. As well as, Agile-SD should have the ability to consider the delayed acknowledgments which needs some modification at the receiver-side.

\section*{Acknowledgments}
This work has been partially supported by the Malaysian Ministry of Education under the Fundamental Research Grant FRGS/02/01/12/1143/FR for financial support.\\

\end{document}